\documentclass[twocolumn]{aastex62}
\usepackage{graphicx}
\usepackage{natbib}
\usepackage{amsmath}	
\usepackage{amssymb}	

\newcommand{\sfro}{\mathrm{SFR}^\mathrm{OUT}}
\newcommand{\sfrt}{\mathrm{SFR}^\mathrm{TOT}}
\newcommand{\fo}{f_{\mathrm{SFR}}}

\newcommand{\msunyr}{M_{\sun}\,\mathrm{yr}^{-1}}
\newcommand{\kms}{\mathrm{km}\,\mathrm{s}^{-1}}

\newcommand{\sigmacl}{\sigma_\mathrm{cl}}

\graphicspath{{./}{PLOTS/}}
\submitjournal{ApJ}

\begin{document}

\title{GASP XXI. Star formation rates in the tails of galaxies undergoing ram-pressure stripping}

\shorttitle{GASP XXI. SFR in the tail of ram-pressure stripped galaxies}
\shortauthors{Gullieuszik et al.}

\correspondingauthor{Marco Gullieuszik}
\email{marco.gullieuszik@inaf.it}

\author[0000-0002-7296-9780]{Marco Gullieuszik}
\affiliation{INAF-Osservatorio Astronomico di Padova, vicolo dell'Osservatorio 5, I-35122 Padova, Italy}

\author[0000-0001-8751-8360]{Bianca M. Poggianti}
\affiliation{INAF-Osservatorio Astronomico di Padova, vicolo dell'Osservatorio 5, I-35122 Padova, Italy}

\author[0000-0003-3255-3139]{Sean L. McGee}
\affiliation{University of Birmingham School of Physics and Astronomy, Edgbaston, Birmingham B15 2TT, UK}

\author[0000-0002-1688-482X]{Alessia Moretti}
\affiliation{INAF-Osservatorio Astronomico di Padova, vicolo dell'Osservatorio 5, I-35122 Padova, Italy}

\author[0000-0003-0980-1499]{Benedetta Vulcani}
\affiliation{INAF-Osservatorio Astronomico di Padova, vicolo dell'Osservatorio 5, I-35122 Padova, Italy}

\author[0000-0002-8710-9206]{Stephanie Tonnesen}
\affiliation{Center for Computational Astrophysics, Flatiron Institute, 162 5th Ave, New York, NY 10010, USA}

\author[0000-0003-2076-6065]{Elke Roediger}
\affiliation{E.A. Milne Centre for Astrophysics, Department of Physics and Mathematics, University of Hull, Hull, HU6 7RX, UK}

\author[0000-0003-2150-1130]{Yara L. Jaff\'e}
\affiliation{Instituto de Fisica y Astronomia, Universidad de Valparaiso, Avda. Gran Bretana 1111 Valparaiso, Chile}

\author[0000-0002-7042-1965]{Jacopo Fritz}
\affiliation{Instituto de Radioastronom\'ia y Astrof\'isica, IRyA, UNAM, Campus Morelia, A.P. 3-72, C.P. 58089, Mexico}

\author[0000-0001-9575-331X]{Andrea Franchetto}
\affiliation{INAF-Osservatorio Astronomico di Padova, vicolo dell'Osservatorio 5, I-35122 Padova, Italy}
\affiliation{Dipartimento di Fisica e Astronomia, Universit\`a di Padova, vicolo dell'Osservatorio 3, I-35122 Padova, Italy}

\author[0000-0002-0838-6580]{Alessandro Omizzolo}
\affiliation{Vatican Observatory, Vatican City State, Vatican City, Italy}
\affiliation{INAF-Osservatorio Astronomico di Padova, vicolo dell'Osservatorio 5, I-35122 Padova, Italy}

\author[0000-0002-4158-6496]{Daniela Bettoni}
\affiliation{INAF-Osservatorio Astronomico di Padova, vicolo dell'Osservatorio 5, I-35122 Padova, Italy}

\author[0000-0002-3585-866X]{Mario Radovich}
\affiliation{INAF-Osservatorio Astronomico di Padova, vicolo dell'Osservatorio 5, I-35122 Padova, Italy}

\author[0000-0001-5840-9835]{Anna Wolter}
\affiliation{INAF-Osservatorio Astronomico di Brera, via Brera 28, I-20121 Milano, Italy}

\begin{abstract}

Using MUSE observations from the GASP survey, we study 54 galaxies undergoing ram-pressure stripping (RPS) spanning a wide range in galaxy mass
and host cluster mass.
We use this rich sample to study how the star formation rate (SFR) in the tails of stripped gas depends on the properties of the galaxy and its host cluster. We show that the interplay between all the parameters involved is complex and that there is not a single, dominant one in shaping the observed amount of SFR.

Hence, we develop a simple analytical approach to describe the mass fraction of stripped gas and the SFR in the tail, as a function of the  cluster velocity dispersion, galaxy  stellar  mass, clustercentric distance and speed in the intracluster medium. Our model provides a good description of the observed gas truncation radius and of the fraction of star-formation rate (SFR) observed in the stripped tails, once we take into account the fact that the star formation efficiency in the tails is a factor $\sim 5$ lower than in the galaxy disc, in agreement with GASP ongoing \ion{H}{1} and CO observations.
We finally estimate the contribution of RPS to the intracluster light (ICL) and find that the average SFR in the tails of ram-pressure stripped gas is $\sim 0.22 \msunyr$ per cluster. By extrapolating this result to evaluate the contribution to the ICL at different epochs, we compute an integrated average value per cluster of $\sim 4 \times 10^9 M_\sun$ of stars formed in the tails of RPS galaxies since $z\sim 1$.
\end{abstract}


\section{Introduction}
Environmental effects play a primary role in galaxy evolution and in
particular in shaping the star formation (SF) history of galaxies in
groups and even more so in clusters \citep[see e.g.][]{bose+2006,
  gugl+2015}.  The capability of a galaxy to form stars crucially
depends on its gas reservoir and, consequently, any process that is able to alter the
content, the dynamics, and the distribution of the gas in a galaxy is
likely to affect its SF history.
Among the external
mechanisms that can potentially impact on the gas content of galaxies
and hence on their SF and evolution, ram-pressure stripping \citep[RPS,][]{gunn+1972} was
proved to be one of the most efficient in clusters of galaxies
\citep{giova+1985,gava1989,kenn+2004,jaff+2015}.  

RPS is the result of the interaction between the galaxy interstellar medium and the hot and
dense intracluster medium (ICM); it
affects only the gas in a galaxy with no direct
effect on its stellar component but it has
dramatic consequences on the formation of new stars.  The most spectacular examples are jellyfish galaxies, that
show tentacles of H$\alpha$
\citep{yosh+2004,sun+2006,yagi+2010,smit+2010,hest+2010,merl+2013,kenn+2014,fuma+2014,ebel+2014,foss+2016,mcpa+2016,gaspXIII}
and/or UV \citep{bois+2012,kenn+2014,geor+2018,pogg+jw100}
emission in the stripped tails.
RPS tails have been also revealed by HI and CO observations \citep{chun+2007,kenn+2004,voll+2009,abra+2011,jach+2014,jach+2017,jach+2019,more+2018,more+2020}.

In the vast majority of the ram-pressure stripped tails that have been studied so far
there is evidence of ongoing star formation (see \citealt{gaspXIII} for a literature review) in agreement with
theoretical predictions of models and
numerical simulations \citep[e.g.][]{kapf+2009,tonn+2012}.
The only known case of a well studied jellyfish galaxy showing an extended H$\alpha$ tail but no reported evidence of ongoing SF is
NGC4569 in the Virgo cluster, which is affected both by ram pressure and
a strong close interaction. \cite{bose+2016}
suggested than mechanisms other than photoionisation,
--such as shocks, heat conduction or magneto-hydrodynamic waves--
are responsible for the gas ionisation.

The first systematic census of jellyfish galaxies were carried out only recently in nearby \citep{pogg+2016} and intermediate-redshift clusters \citep{ebel+2014,mcpa+2016}.
The \cite{pogg+2016} sample was used to select the targets
for the GAs Stripping Phenomena in galaxies \citep[GASP,][]{gaspI} survey,
which is based on an ESO Large Programme
that was awarded 120 h observing time with the MUSE IFU at the VLT to
observe 114 galaxies at $z = 0.04\mbox{--}0.07$ in galaxy clusters and in the field.
GASP MUSE data are complemented by ongoing
observing campaigns with JVLA, APEX and ALMA to probe the cold atomic and molecular gas component.  We are also collecting near- and far-UV imaging with UVIT on-board ASTROSAT to search for UV tails tracing SF regions.
MUSE observations have demonstrated that SF is ubiquitous in the tails of GASP jellyfish galaxies \citep{gaspXIII}, in agreement with the detection
of UV light \citep{geor+2018}; CO observations proved that
the tail's stellar component
formed in-situ from large
amounts of molecular gas detected well outside of the galaxy disks
\citep{gaspX, more+2020}.

Jellyfish galaxies offer the unique opportunity to
study the SF process
in the peculiar environment of the gas-dominated tails, in the absence of an underlying galaxy disk.
Moreover, 
the SF processes in the tails could be influenced by thermal conduction from
the hot ICM, which might heat the 
gas and therefore prevent it from collapsing into clouds.
The first systematic study of the properties of SF regions in the tails of 16 jellyfish galaxies, based on GASP data, was presented by \cite{gaspXIII}. 
In-situ SF was found to be ubiquitous in the tails of RPS galaxies, taking place in large and massive clumps with a median stellar mass up to $3\times 10^7 M_\sun$; these clumps could therefore play a role in the formation of the population of ultra-compact dwarf galaxies, globular clusters, and dwarf spheroidal galaxies in clusters.

Studying SF in the tails of jellyfish galaxies and assessing the SFR in a statistically significant sample is fundamental to understand galaxy evolution as well as the role of RPS in building-up the stellar component of the ICM and the intracluster light (ICL). There is in fact still some tension between the conclusions of different works about the relative contribution of e.g.
disruption of dwarf galaxies, violent mergers, tidal stripping of stars and
in situ formation from stripped gas \citep[see][and references therein]{gial+2014,adam+2016,mont+2018,cont+2018,dema+2018}.

In this paper we use the complete sample of GASP galaxies in clusters to measure the 
ongoing star formation in the tails and in the galaxy main body to investigate how the amount and fraction of SFR in the tails depend on the properties of the galaxy and of the host cluster, and on the orbital properties of the galaxy in the host cluster, 
i.e. on the projected position and velocity of the galaxy relative to the cluster.
The observed SFR in the tails is then used to estimate the integrated contribution of RPS to the ICL.

This paper is organised as follow:
in Sect. \ref{sec:data} we present our data and  
in Sect. \ref{sec:analysis} we describe our measurements methods and analysis;
in Sect. \ref{sec:obsres} we presents our observational results;
in Sect. \ref{sec:model} we propose a simple analytical approach to
estimate the fraction of SFR in the tails and we compare the results with our observations;
in Sect. \ref{sec:icl} we use our data to estimate the total contribution to the ICL
due to ram-pressure stripping;
in Sect. \ref{sec:conclusions} we summarise our work and conclusions.

This paper adopts the \cite{chab+2003} initial mass function and the standard concordance cosmology:
$H_0=70$ $\mathrm{km}\,{{\rm{s}}}^{-1}\,{\mathrm{Mpc}}^{-1},{{\rm{\Omega }}}_{M}\,=0.3$, ${{\rm{\Omega }}}_{{\rm{\Lambda }}}=0.7$

\section{Data}\label{sec:data}
GASP observations were carried out between October 2015 and April 2018 in service mode with the 
Multi Unit Spectroscopic Explorer (MUSE) integral-field spectrograph \citep{baco+2010}
mounted at the Nasmyth focus of the UT4 VLT, at Cerro Paranal in Chile. 
The MUSE spectral range, between 4500 and 9300 \AA, is sampled at 1.25 \AA\, pixel$^{-1}$, with a spectral resolution of $\sim2.6$ \AA. 
The $1^\prime \times 1^\prime$ field of view is sampled at $0.2$ arcsec\, pixel$^{-1}$; each datacube therefore consists of $\sim10^5$ spectra.
The MUSE FoV is large enough to completely cover 109 GASP targets;
the remaining 5 targets (namely JO60, JO194, JO200, JO201, JO204, and JO206) were observed combining two pointings.

Raw data were reduced using the latest ESO MUSE pipeline available when observations were taken. The reduction procedure 
and the methods used for GASP data analysis 
are described in detail in \cite{gaspI}.
The sky-subtracted, flux-calibrated datacubes  are corrected for Galactic extinction using the \cite{schl+2011} reddening map and the \cite{card+1989} extinction law.
As a first step, to increase the signal-to-noise ratio in the
low surface brightness regions, we applied a 5-pixel-wide boxcar filter 
in the spatial directions, replacing the value of each spaxel, at each wavelength, with the average value of the $5\times5$ neighbouring spaxels.
The kinematic of the stellar component was derived using
the pPXF code \citep{capp+2004} and the spatially resolved properties of the stellar populations were obtained using our spectro-photometric fitting code SINOPSIS \citep{gaspIII}; 
the emission-only spectra of the gas component
were computed by subtracting to the observed spectra the best-fit stellar model obtained with SINOPSIS to the datacubes corrected for extinction from our Galaxy.
Gas emission line fluxes, velocities and velocity
dispersions with associated errors were then computed using 
KUBEVIZ \citep{foss+2016}.
As a last step, we corrected the absorption-corrected line emission fluxes
for the intrinsic extinction using the Balmer decrement, assuming a value
H$\alpha$/H$\beta$ = 2.86 and the \cite{card+1989} extinction law.

The 114 GASP galaxies include:
(i) 64 cluster galaxies selected from the  \citet{pogg+2016} catalogue of candidate gas stripping galaxies;
(ii) 12 control sample cluster galaxies from the WINGS \citep{fasa+2006} and OmegaWINGS survey \citep{gull+2015};
(iii) 38 galaxies in low density environments (groups and filaments):
30 stripping candidates from the \citet{pogg+2016} catalogue
and 8 control sample galaxies from the Padova Millennium Galaxy and Group Catalog 
(PM2GC, \citealt{calv+2011}
Colour images and H$\alpha$ emission maps of all 114 galaxies are available on a webpage at
\url{http://web.oapd.inaf.it/gasp/gasp_atlas}.
Redshift measurements obtained from MUSE observations showed that five of the 64 stripping candidates in clusters are actually
non-members (nine of the GASP candidates in \citealt{pogg+2016} had no redshift measurement). Other five galaxies are found to have close companions and therefore they are likely merging/tidally interacting systems. 
For this paper we will consider the remaining 54 non-interacting gas stripping candidates in clusters which are listed in 
Table~\ref{tab:tab}. 
The cluster redshift and velocity dispersion in Table~\ref{tab:tab} are taken from \cite{bivi+2017} and \cite{more+2017};
the virial radius $R_{200}$ is taken from \cite{bivi+2017} when available; otherwise
they
are derived from the observed line of sight velocity dispersion by using Eq.~1 in \cite{muna+2013} and the relation
\begin{equation}
R_{200}=\left(\frac{G~M_{200}} {100~H_z^2}\right)^{1/3}
\end{equation}
where $H_z$ is the Hubble constant at the redshift of the cluster.

\section{analysis}\label{sec:analysis}
\subsection{Galaxy boundary definition}\label{sec:maskdef}

This paper is aimed at quantifying the amount of SFR in the tails of stripped gas as a function of galaxy and cluster properties, and galaxy orbital histories within the cluster.  To define the tails, we would ideally need to assess the location of gas that is not gravitationally bound to the galaxy main body, which is clearly operatively not feasible.

\begin{figure*}[!ht]
\centering
\includegraphics[width=.85\textwidth]{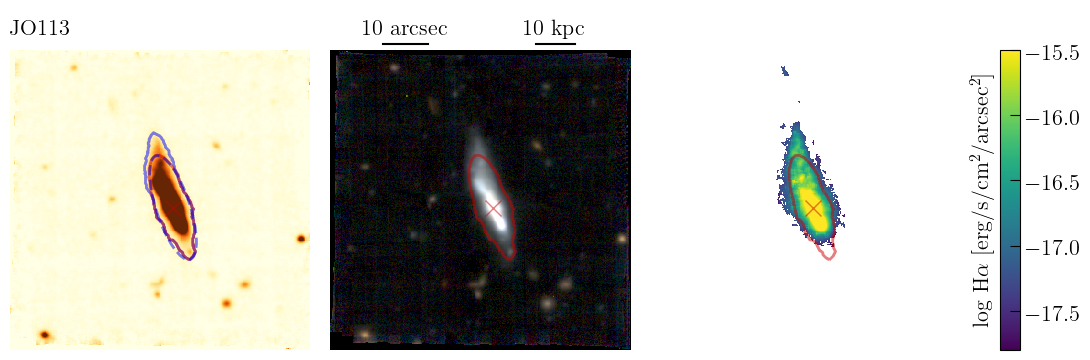}\\
\includegraphics[width=.85\textwidth]{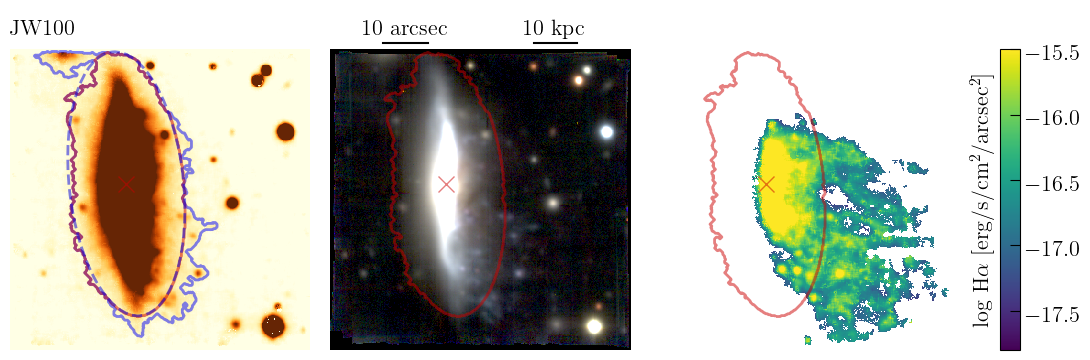}
\includegraphics[width=.85\textwidth]{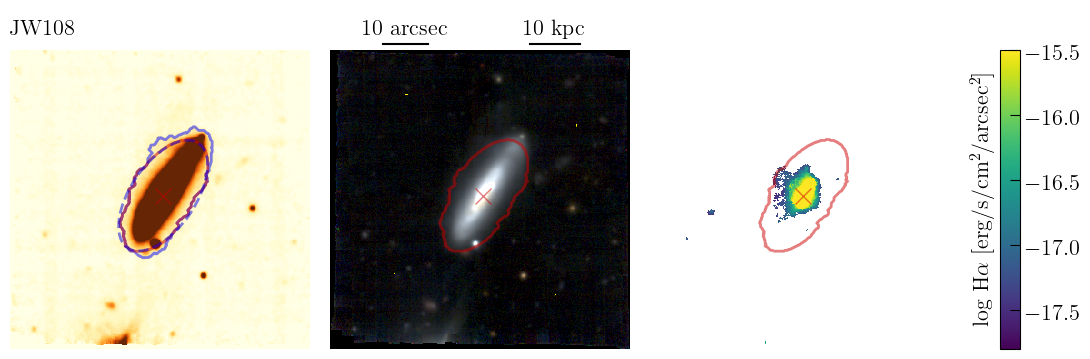}\\
\caption{From left to right: (i) the continuum emission in the H$\alpha$ region; (ii) an RGB image obtained from $g-$, $r-$, and $i-$band images derived from the MUSE datacube; ; (iii) the H$\alpha$ emission of 
three galaxies at different stripping stages: JO113 (moderate stripping), JW100 (jellyfish galaxy), and JW108 (truncated disc or post stripping).
The blue line is the isophote corresponding to a continuum surface brightness 1$\sigma$ above the background level; the dashed line is the ellipse that better describes the isophote on the undisturbed side of the galaxy and the red line is the resulting line that we used to define the galaxy main body. The same figure for all galaxies is available at \url{http://web.oapd.inaf.it/gasp/inandout}.}
\label{fig:inoutmap}
\end{figure*}

We developed a procedure to define the 
galaxy boundary and estimate a conservative lower-limit to the amount of stripped material by using the 
continuum map obtained by the KUBEVIZ model of the H$\alpha$+[\ion{N}{2}] lineset,
to probe the stellar galactic component. First of all we defined the centre of the galaxy as the centroid of the brightest central region
in the continuum map. The resulting centre positions are
listed in Table~\ref{tab:tab}.
We then considered the faintest visible stellar isophote, which is
defined as the one corresponding to a surface brightness $1\sigma$ above the sky background level. 
For galaxies undergoing ram-pressure stripping, this isophote does not have elliptical symmetry because of the emission from stars born in the stripped tail and of the (minor) contribution from the gas continuum emission. For this reason, we fit an ellipse to the undisturbed side of the isophote; this ellipse was used to replace the isophote on the disturbed side. The resulting contour defines a mask that we used to discriminate the galaxy main body and the ram-pressure stripped tail. 
In the following, we will refer to 
the regions within this mask
as the galaxy main body, and to the regions outside the mask as tails.
This definition of the galaxy main body and tail was already exploited by \citet{gaspXIII} and \citet{gaspXIV}.
Examples to illustrate the definition of the mask for three galaxies at different stripping stages --the same three prototypical galaxies 
used by \cite{gaspIX}--
are shown in Fig.~\ref{fig:inoutmap}; the same figures for all galaxies are available at \url{http://web.oapd.inaf.it/gasp/inandout}.

\subsection{Stellar mass and star formation rates} \label{sec:mass_sfr}

As in \cite{gaspXIV},  we define the stellar mass of the galaxy main body $M_\star$ as the sum of the stellar mass computed with SINOPSIS for each spaxel within the galaxy main body mask. The resulting values are listed in Table~\ref{tab:tab} and shown in Fig.~\ref{fig:histos}; they range between $6\times 10^8$ and $3\times 10^{11} M_{\sun}$. We note that besides SOS 114372 \citep[][which is the GASP galaxy JO147]{merl+2013} all other jellyfish galaxies studied in the literature before GASP have stellar masses below $3\times 10^{10} M_{\sun}$ (see the literature review in \citealt{gaspXIII}).

To investigate the gas ionisation mechanism we used the standard BPT diagrams \citep{bpt}, based on the
[\ion{O}{3}]5007/H$\beta$ vs [\ion{N}{2}]6583/H$\alpha$ line ratios.
Following \cite{gaspI}, 
we adopted the classification scheme based on the results of
\cite{kewl+2001}, \cite{kauf+2003}, and \cite{shar+2010}
to separate regions with AGN- and LINER-like from star-forming and composite (star-forming+LINER/AGN) regions.
In most galaxies the tails are ionised mainly by massive young stars; 
extended regions with AGN-like emission are observed in JO135 and JO204
which are likely due to the ionisation cone of the central AGN \citep{gaspIV,gaspVI,gaspXIX}.

The SFR was computed from the H$\alpha$ flux corrected for
stellar and dust absorption excluding the regions classified as AGN or LINERS and adopting Kennicutt's relation for a \cite{chab+2003} IMF:
\begin{equation}
\mathrm{SFR}=4.6\times 10^{-42} L_{\mathrm{H}\alpha}.
\end{equation}
The total SFR and the SFR in the tails (hereafter $\sfrt$ and $\sfro$, respectively) for all galaxies considered in this paper are listed
in Table~\ref{tab:tab}.

\begin{deluxetable*}{lrrlrrrrrrrc}
\tablecaption{GASP galaxies used in this work}
\tablehead{
\colhead{ID} & \colhead{RA} & \colhead{DEC} & \colhead{cluster} & \colhead{$\sigmacl$} &
\colhead{log$M_\star$} &
\colhead{$z$} & \colhead{$\sfrt$} & \colhead{$\sfro$}&
\colhead{$r_\mathrm{cl}/R_{200}$} &
\colhead{$v/\sigmacl$} &
\colhead{JC}\\
&\colhead{(J2000)}&\colhead{(J2000)}&&
\colhead{(km/s)}&
&&
\colhead{($\msunyr$)}&
\colhead{($\msunyr$)}&
&&}
\colnumbers
\startdata
JO5 & 10:41:20.38 & -08:53:45.6 & A1069 & 542 & 10.27 & 0.0648 & 1.350 & 0.031 & 1.72 & 0.26 & 3 \\
JO10 & 00:57:41.61 & -01:18:44.0 & A119 & 952 & 10.76 & 0.0471 & 3.083 & 0.000 & 0.50 & 0.81 & 1 \\
JO13 & 00:55:39.68 & -00:52:36.0 & A119 & 952 & 9.82 & 0.0479 & 1.546 & 0.002 & 0.57 & 1.05 & 4 \\
JO17 & 01:08:35.33 & 01:56:37.0 & A147 & 387 & 10.16 & 0.0451 & 0.847 & 0.000 & 1.05 & 0.28 & 1 \\
JO23 & 01:08:08.10 & -15:30:41.8 & A151 & 771 & 9.67 & 0.0551 & 0.298 & 0.000 & 0.45 & 0.67 & 1 \\
\enddata
\tablecomments{This table is published in its entirety in the machine-readable format.
      A portion is shown here for guidance regarding its form and content.}
\tablecomments{Columns are:
1) GASP ID number from \citet{pogg+2016};
2) and 3) Equatorial coordinate of the galaxy centre;
4) host cluster;
5) velocity dispersion of the host cluster;
6) logarithm of the galaxy stellar mass  (in solar masses);
7) galaxy redshift;
8) total SFR;
9) SFR in the tails;
10) projected distance from the cluster centre in units of $R_{200}$;
11) line of sight velocity of the galaxy with respect to the cluster mean in units of the cluster velocity dispersion;
12) Jellyfish Class (JC) from \citet{pogg+2016}; 
\label{tab:tab}
}
\end{deluxetable*}

\section{Observational Results}\label{sec:obsres}

\begin{figure}
\centering
\includegraphics[width=\columnwidth]{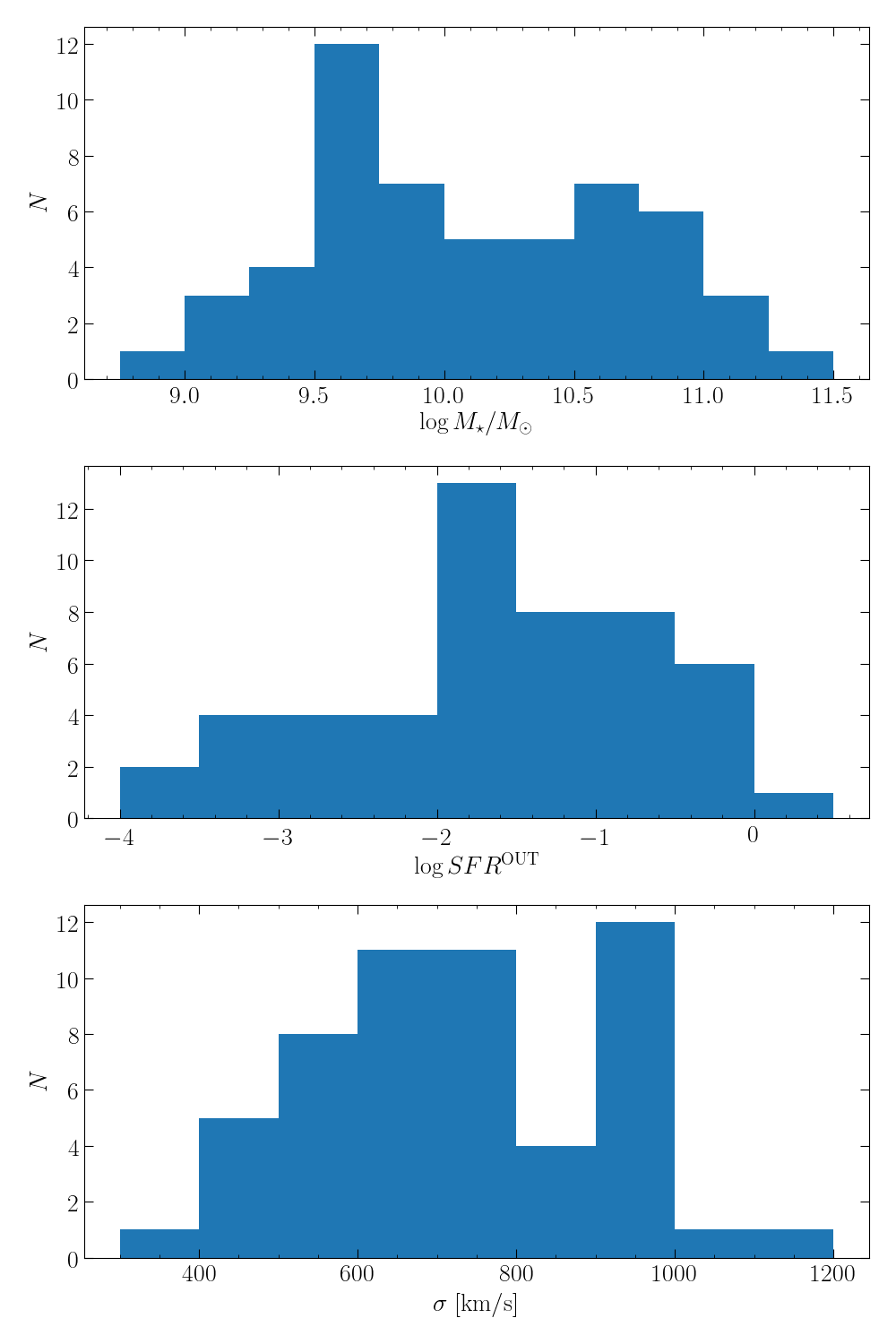}
\caption{Distribution of the stellar mass (upper panel), SFR in the tails (central panel) and velocity dispersion of the host cluster (lower panel) for the sample galaxies.}
\label{fig:histos}
\end{figure}

Figure \ref{fig:histos} shows the distributions of galaxy stellar masses, measured 
$\sfro$, and host cluster velocity dispersion. We are sampling galaxies with a wide range of stellar masses, from less than
$10^9$ to $10^{11.5} M_{\sun}$, hosted in low- and high-mass clusters, with a velocity dispersion from 
400 to more than 1000 km s$^{-1}$.
The SFRs  we measured in the tails show a wide variation, reaching values up to more than $1 \msunyr$.
The decreasing number of galaxies with low $\sfro$ is likely strongly affected by incompleteness, due to observational effects and/or selection biases.

\begin{figure}
\centering
\includegraphics[width=\columnwidth]{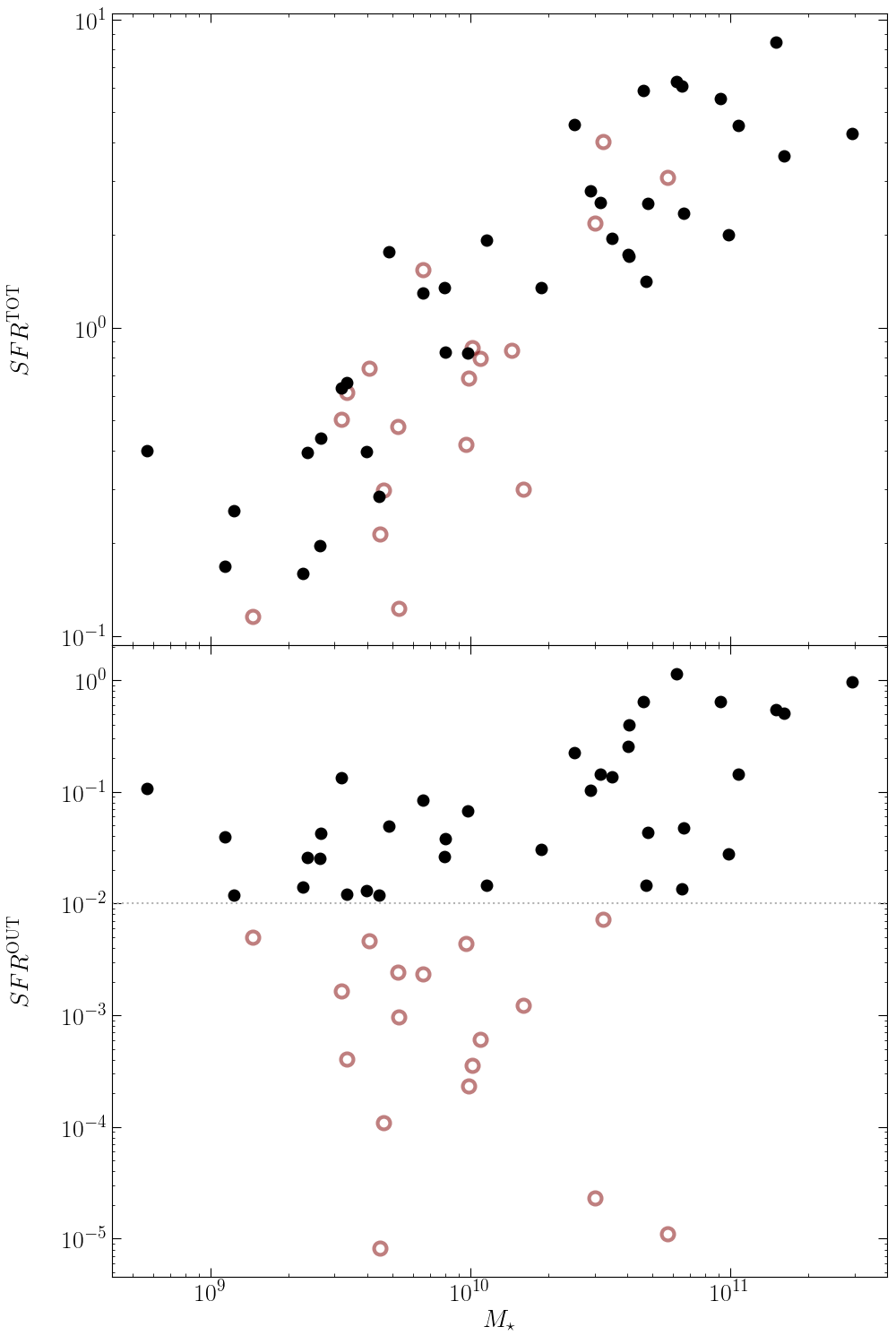}\\
\caption{Total SFR (upper panel) and SFR in the tails (lower panel) as a function of the galaxy stellar mass.
Red circles shows galaxies with SFR in the tail smaller than 
$10^{-2} \msunyr$.}
\label{fig:sfrmass}
\end{figure}

The correlation between $\sfrt$ and galaxy stellar mass
is shown in the upper panel in Fig.~\ref{fig:sfrmass};
a detailed analysis of this correlation is  
presented in \cite{gaspXIV}, which demonstrates a statistically significant enhancement of the SFR in both the discs and the tails of 
GASP ram-pressure stripped galaxies compared to undisturbed galaxies. The lower panel  in Fig.~\ref{fig:sfrmass}
shows the relation between the SFR in the tails and the galaxy stellar mass.
We define as galaxies with a significant SFR in the stripped tails
 those with a SFR outside the mask defined in the previous section larger than
$10^{-2} \msunyr$; these are shown as black filled symbols in Fig.~\ref{fig:sfrmass} and do not show a 
clear and well defined correlation between the SFR in the tail and the stellar mass. 
All galaxies with $M<10^{10}M_{\sun}$ are forming stars in the tails
at a rate $\lesssim 0.1 \msunyr$, while only among the most massive galaxies we observe SFR in the tails above this value and up to $1.6 M_{\sun} \mathrm{yr}^{-1}$.
In addition
we found galaxies with low SFR in the tail (open red circles in the lower panel of Fig.~\ref{fig:sfrmass}) at all masses below $10^{11} M_{\sun}$. 

The value of the observed $\sfro$ is expected to depend 
on many different factors: 
(i) the total amount of gas available for forming stars;
(ii) the efficiency of RPS which, in turn depends on the strength of the ram pressure and on the galaxy anchoring force (see Sect. \ref{sec:model});
(iii)
the star formation efficiency in the tails, which, in principle, can be different from the one in the galaxy main body.
In the following, we use 
large number of ram-pressure stripped galaxies in the GASP sample 
to search for general trends.

\begin{figure}
\centering
\includegraphics[width=\columnwidth]{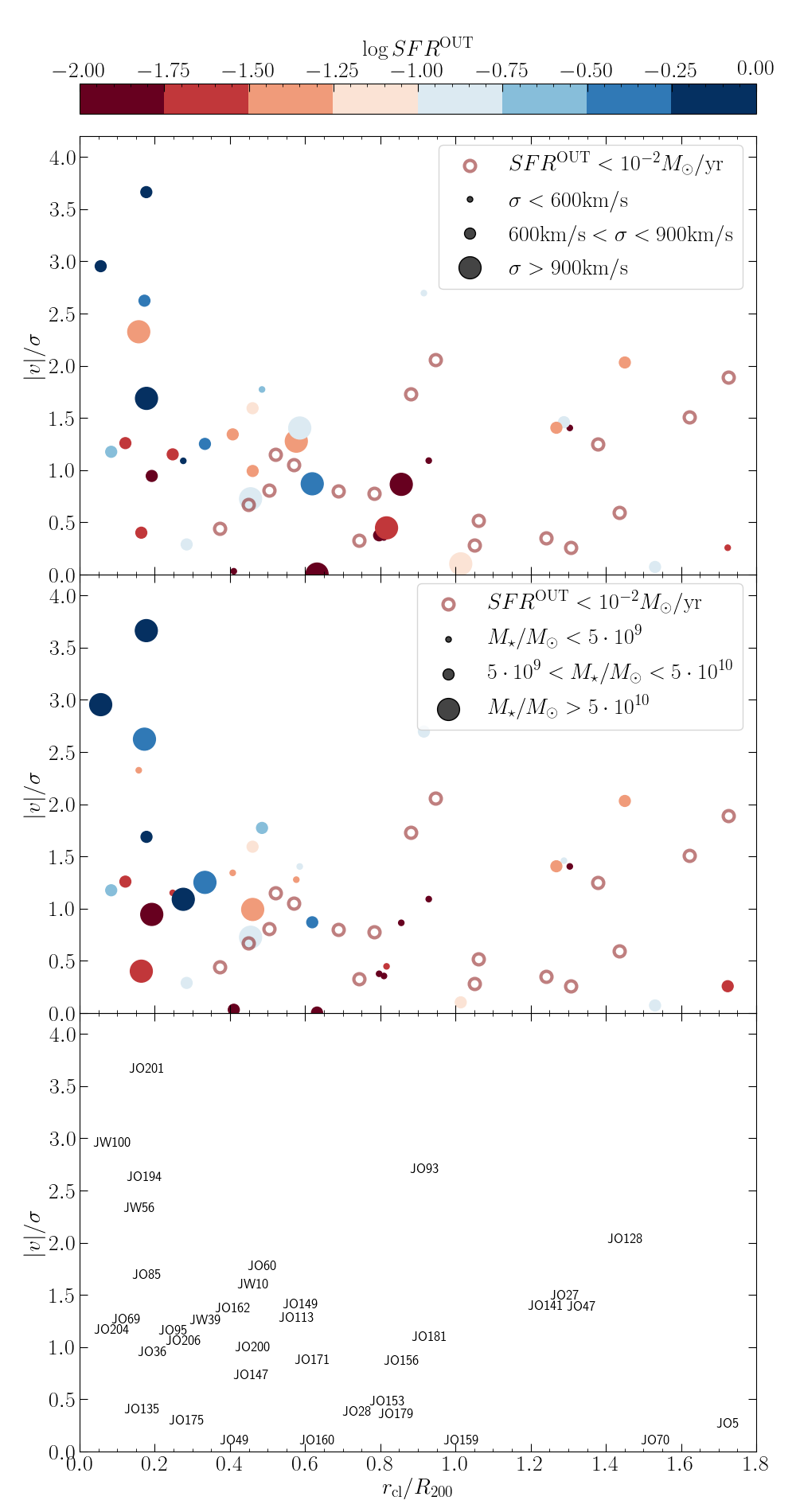}\\
\caption{In the two upper panels we plot the phase-space diagram for
galaxies with marginal or no star formation in the tails ($\sfro<10^{-2} M_{\sun}/\textrm{yr}$) as red circles. All other galaxies 
are colour coded according to the SFR in the tail, as shown in the upper bar. In the upper panel, the point size indicates the host cluster velocity dispersion, in the lower panel the galaxy stellar mass. 
The bottom panel shows the names of the galaxies with $\sfro\geq 10^{-2} M_{\sun}/\textrm{yr}$ at the points positions.}
\label{fig:psd}
\end{figure}

The position versus velocity phase-space diagram is an extremely useful tool
to investigate environmental effects on the evolution of galaxies in clusters and it has been effectively used
by \cite{gaspIX} to correlate the stripping stage of GASP galaxies
with their orbital histories.
The projected phase-space diagram of all our target galaxies is shown in Fig.~\ref{fig:psd};
all galaxies with marginal ($< 10^{-2} \rm M_{\odot} yr^{-1}$, empty red circles) SFR in the tails are
found at relatively large projected distances 
($r_{\textrm{cl}}>0.3 R_{200}$).
Among these, those at high speed are most likely 
being accreted (therefore have not been strongly stripped yet), while those at lower speed statistically have spent already more time in the cluster ($>2$Gyr)
and probably have little gas left.
However, projection effects can blend these two populations.
Instead, galaxies with a conspicuous SFR in the tails
are preferentially found in the inner cluster regions ($r_{cl}/R_{200} < 0.6$) moving at high speed ($|v|>\sigma$) in the ICM (see the colourbar), which suggests they are on first infall into the cluster, on preferentially radial orbits. 
Figure~\ref{fig:psd} shows 
that in the most massive clusters (those with large velocity dispersion shown with the largest symbols in the upper panel in Fig.~\ref{fig:psd})
it is possible to find galaxies with a significant 
amount of SF in the tail also
at relatively large distance from the cluster centre (up to $\sim 0.5-0.6 R_{200}$
and/or moving at not extreme velocities ($v/\sigmacl \lesssim 1.5$).

Galaxies moving at high speed ($|v|>2\sigmacl$) in the innermost cluster regions ($r_{\mathrm{cl}}<0.5 R_{200}$) --the region of the phase-space diagram where the maximum effect of RPS is expected-- 
that have a large SFR in the tails ($\log \sfro >-0.25$)
are all massive and are hosted in low mass clusters (large and small symbols in the lower and upper panel of Fig.~\ref{fig:psd}, respectively). In these galaxies, therefore, the internal anchoring force is stronger due to the gravitational potential of the galaxy itself, and the expected ram-pressure is not so extreme due to the low-mass cluster environment. These galaxies should be the ones able to retain a significant fraction of their gas all the way until they reach
the central regions of the cluster. If they were less massive, or within a more massive cluster, they would have been totally stripped before they reached short clustercentric distances.
Their value of $\sfro$ needs to be explained taking into account, therefore, all the parameters cited above.

Three of the four galaxies with $\sfro < 10^{-2} \msunyr$
that are found within $r_{\mathrm{cl}}<0.5 R_{200}$
have truncated H$\alpha$ discs, namely JO10, JO23, and JW108. These galaxies
likely developed tails at some point in the past that are now completely stripped. 
The only other GASP galaxy with a truncated disc is JO36. This galaxy has a 
$\sfro=0.014 \msunyr$, barely above the threshold we adopted, and indeed it
lies in the same region of Fig.~\ref{fig:psd} as the other truncated discs (see Table~\ref{tab:tab})
and has a similarly low $\sfro$.
We note that JO36 is likely undergoing also a gravitational interaction by a fly-by or a close encounter with another galaxy in the cluster which may have stripped part of the gas and hence increased the $\sfro$ \citep{gaspIII}.
Although its very difficult to isolate back-splashing galaxies in the phase-space due to projection effects, these truncated-disc galaxies are good candidates (see also \citealt{yoon+2017})

\begin{figure}
\centering
\includegraphics[width=\columnwidth]{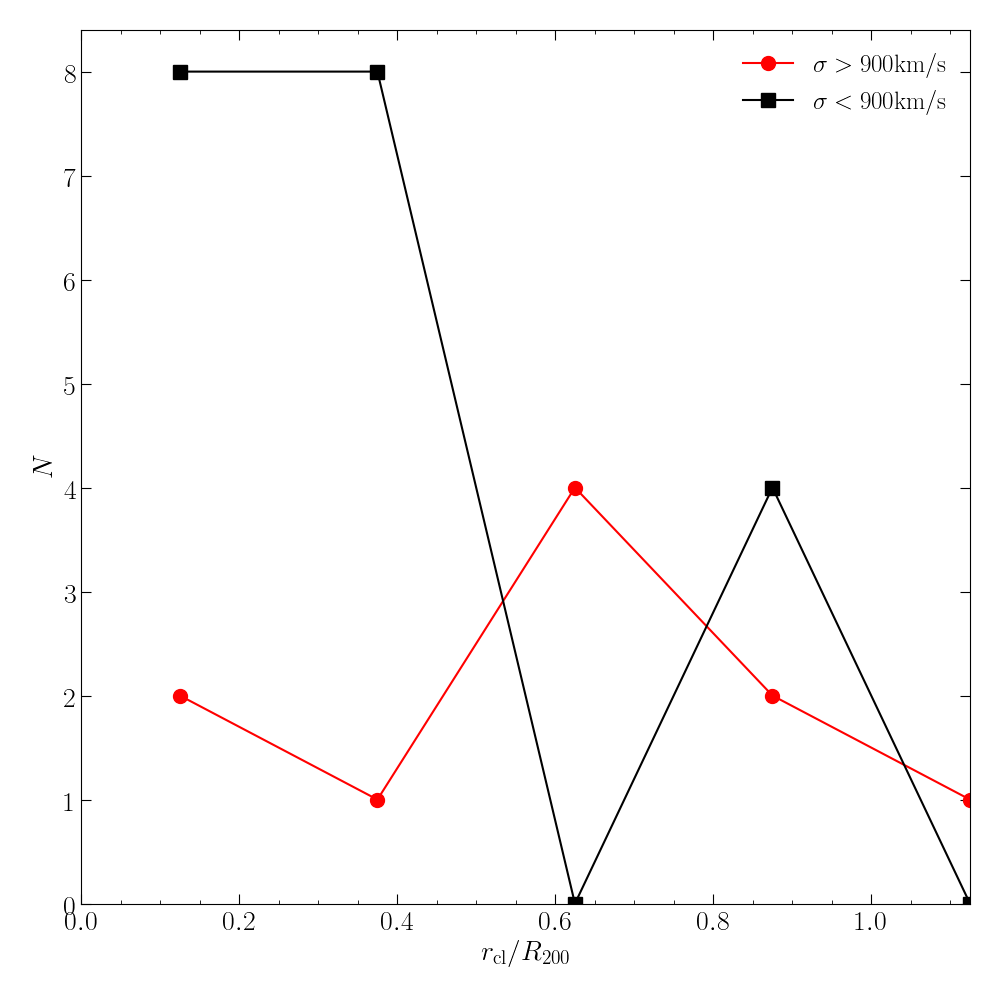}
\caption{Radial distribution of galaxies with $\sfro>10^{-2} \msunyr$.}
\label{fig:radial_massive_cl}
\end{figure}

Only a few galaxies with a significant $\sfro$ ($>10^{-2}\msunyr$) are hosted in the central regions ($r_{\mathrm{cl}}<0.5R_{200})$ of very massive clusters while the majority of them belong to intermediate and low-mass environments.
A direct evidence of this is shown in Fig.~\ref{fig:radial_massive_cl}:
while 
the radial distribution of galaxies with $\sfro>10^{-2} M_{\sun}$ hosted in clusters
with $\sigmacl < 900 \mathrm{km} \, \mathrm{s}^{-1}$ increases toward the central
regions, the distribution of galaxies in clusters with 
$\sigmacl > 900 \mathrm{km} \, \mathrm{s}^{-1}$ is flat
at $r_{\mathrm{cl}}\lesssim 0.6 R_{200}$. Since the amount of SF in the tails is intimately linked to the gas stripping efficiency, this may be an indication that RPS occurs preferentially at intermediate clustercentric 
radii in massive clusters and at lower radii in intermediate and low-mass clusters.
One of the 2 galaxies in the inner regions of massive clusters is JO85; this is a lopsided galaxy that is very likely undergoing nearly edge-on stripping,
that is substantially less efficient than face-on stripping (see for example the simulations in \citealt{roed+2014}); this would explain why a substantial fraction of the gas was not stripped during the infall.

\begin{figure}[!ht]
\centering
\includegraphics[width=0.95\columnwidth]{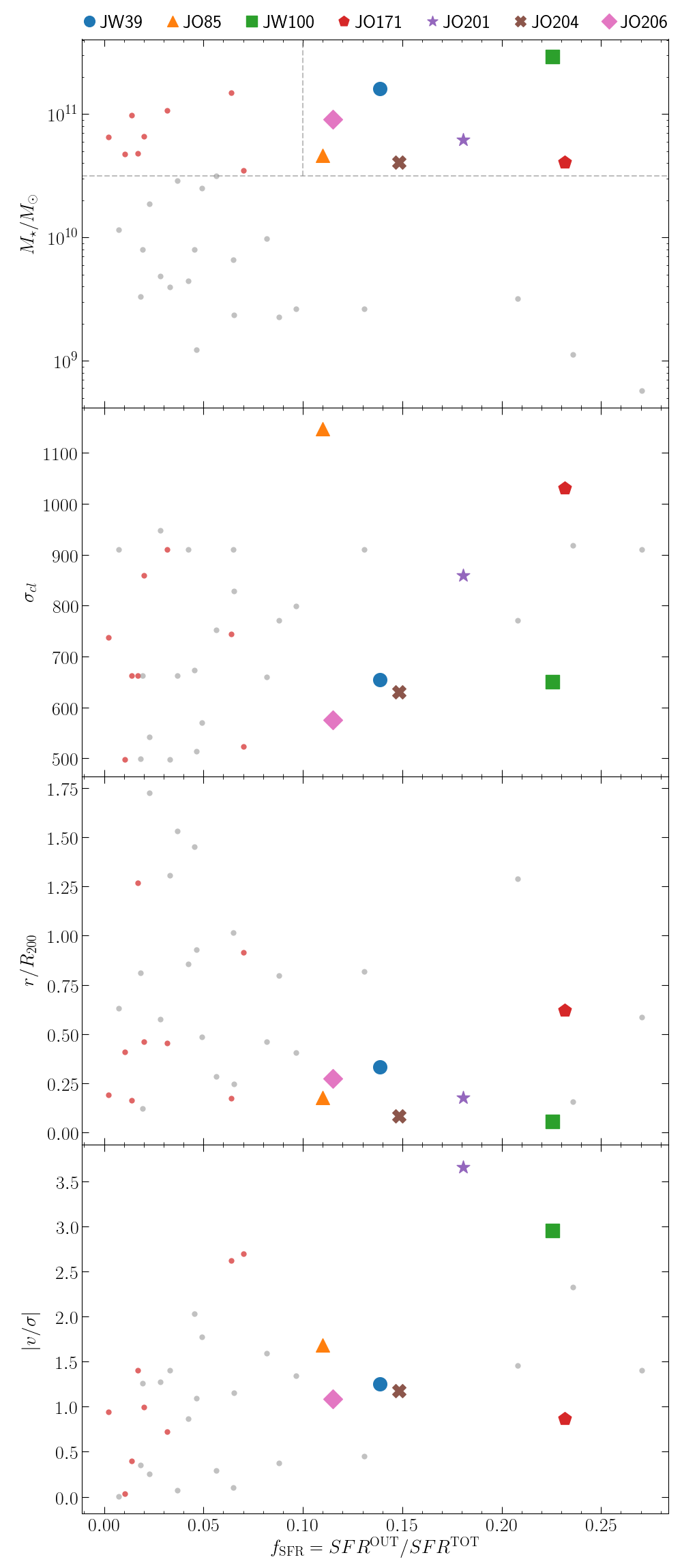}
\caption{Fraction of SFR in the tails as a function of --from top to bottom panels-- the disc stellar mass, the host cluster velocity dispersion, the clustercentric distance and the galaxy radial velocity. Small grey symbols are galaxies with $M_\star<10^{10.5}M_{\sun}$; red symbols are for galaxies more massive than this limit and a fraction of SFR in the tail smaller than 10\%. Large symbols are massive galaxies with more then 10\% of SFR in the tail; names of individual galaxies are given in the legend at the top.}
\label{fig:sfratiomassive}
\end{figure}

The upper panel in Fig.~\ref{fig:sfratiomassive} shows the fraction of SFR in the tail $\sfro/\sfrt$ ($\fo$ thereinafter)
as a function of the stellar mass (only for galaxies with $\sfro>10^{-2} M_{\sun} \mathrm{yr}^{-1}$).
As already pointed out, the gravitational potential is much stronger in high-mass galaxies and consequently the anchoring force is stronger than in low-mass galaxies. We see that among the most massive galaxies ($\log M_{\star}/M_{\sun}>10.5$) there are 7 galaxies with a substantial fraction of
SFR in the tail, $\fo>10\%$.
The ram-pressure acting on these galaxies should hence be particularly intense to overcome the anchoring force and strip the gas.

In the lower panels in Fig.~\ref{fig:sfratiomassive} we show $\fo$ as a function of the main observable quantities regulating the ram-pressure intensity, namely the cluster velocity dispersion (a proxy for the cluster mass, thus related to the ICM density), the clustercentric distance and the peculiar velocity within the ICM normalised by the cluster velocity dispersion. Let us consider the 7 massive galaxies with very high $\fo$ values. JW100 is the most massive galaxy in the GASP sample and it is hosted in a relatively low-mass cluster, but it is moving at very high speed and it is very close to the cluster centre, where
the ICM is denser. JO85 and JO171 are hosted in two of the most massive clusters ($\sigmacl>1000\kms$), while JO201 is moving at extremely high speed in the ICM of a relatively massive cluster.
Finally, JO204, JO206 and JW39 are very close to the centre of their host cluster centre and moving at quite high speed.
All these seven galaxies are in exceptional conditions regarding at least one the main parameters regulating the ram-pressure intensity. 
We note that, besides JO201, all other six galaxies have a tail of stripped gas
with a long extension on the plane of the sky, indicating that the component on the plane of the sky of the velocity of the galaxies in the ICM is dominant with respect to the one along the line of sight; the measured radial velocity is therefore a lower limit of their actual speed.
Five of these galaxies are studied in detail in dedicated papers (
JW100, \citealt{pogg+jw100}; 
JO171, \citealt{gaspV};
JO201, \citealt{gaspII,gaspXV};
JO204, \citealt{gaspIV};
JO206, \citealt{gaspI}). 

To summarise, in this section we have described the star formation occurring in the tails of ram-pressure stripped galaxies in terms
of both the properties of the galaxies and of the host cluster. We found general trends, but the interplay between all the parameters involved in defining the actual value of $\sfro$ is complex and all of them must be taken into account. 
To better understand this scenario we developed an analytical model based on \citet{gunn+1972} prescriptions aimed at providing an estimate of $\sfro$ as a function of a limited number of relatively easily measurable quantities.

\section{The analytical model}\label{sec:model}

This section presents our analytical approach based on  \citet{gunn+1972} prescriptions to evaluate the fraction of star formation in the tail of ram-pressure stripped galaxies and following the work presented in  \cite{gaspIX}, with a formulation similar to \citet{smit+2012} and \citet{ower+2019}.

The ram-pressure on a galaxy moving at a speed $v$ in an ICM with density $\rho$ is
\begin{equation}
P_\mathrm{RAM}=\rho v^2
\end{equation}

The gas within a galaxy will be stripped when $P_\mathrm{RAM}$ overcomes the galaxy's anchoring force $\Pi$ which can be modelled assuming the form:  
\begin{equation}
\Pi=2\pi G \Sigma_\star \Sigma_g 
\end{equation}
where $G$ is the gravitational constant, $\Sigma_g$ and $\Sigma_\star$ are the surface density profiles of the gas  and stellar discs, respectively.
We assumed an exponential profile for both of them: 
\begin{eqnarray}
\Sigma_\star=&\frac{M_{d,\star}}{2\pi R_{d,\star}^2} \, e^{-r/R_{d,\star}}\\
\Sigma_g=&\frac{M_{d,g}}{2\pi R_{d,g}^2} \, e^{-r/R_{d,g}}
\end{eqnarray}
where $M_{d,g}$ and $M_{d,\star}$ are the mass, and $R_{d,g}$ and $R_{d,\star}$ are the scale lengths of the gas and stellar discs.

We also assumed that:
\begin{itemize}
\item
galaxies are disc dominated, we therefore defined $M_\star \equiv
M_{d,\star}$;
\item
the  gas-to-stellar scale length ratio is 1.7, as in \cite{gaspIX};
this corresponds to the ratio between the \ion{H}{1} and optical radius of
non-\ion{H}{1}-deficient galaxies in Virgo cluster \citep{caya+1994}.
In the following, the stellar disc scale length will be referred to as
$R_d$\footnote{\citet{bigi+2012} found that the total (\ion{H}{1}+H$_2$) gas scale length is $0.48 \pm 0.04 R_{25}$;
using the scaling factor $R 25/R_d = 4.6 \pm 0.8$ from \cite{lero+2008},
the \ion{H}{1}-based gas-to-stellar scale-length ratio we assumed in this paper
is compatible with the \citet{bigi+2012} result within uncertainties.}.
\end{itemize}

We define the truncation radius $r_t$ as the distance from the galaxy centre where
$\Pi=P_\mathrm{RAM}$. At radii larger than $r_t$ ram pressure overcomes the anchoring force
and the gas is stripped.
If we call $f_\mathrm{gas}=M_{d,g}/M_\star$
the gas mass fraction,
we have:

\begin{equation}
\begin{split}
\rho v^2  = & 2\pi G \cdot 
\frac{M_\star}{2\pi R_d^2}e^{-r_t/R_d} \cdot 
\frac{f_\mathrm{gas} M_\star}{2\pi\, 1.7^2 R_d^2} e^{-r_t/(1.7 R_d)} \\
= & \frac{f_\mathrm{gas} GM_\star^2}{2\pi 1.7^2 R_d^4} 
e^{-2.7r_t/(1.7R_d)}
\end{split}
\end{equation}

Taking the logarithm of both sides of this equation yields

\begin{equation}
\ln(\rho \, v^2) = \ln \frac{f_\mathrm{gas} GM_\star^2}{2\pi 1.7^2 R_d^4} -\frac{2.7r_t}{1.7R_d}
\end{equation}

and hence we obtain

\begin{equation}\label{eq:rtrd}
\frac{r_t}{R_d}= \frac{1.7}{2.7} \left[
\ln \frac{f_\mathrm{gas} GM_\star^2}{2\pi 1.7^2 R_d^4}  -
\ln(\rho v^2) 
\right]
\end{equation}

The fraction of remaining gas mass in the galaxy can be calculated by integrating the mass distribution
of an exponential disc assuming that all gas outside
the truncation disc is stripped and lost. If we call 
$M_g$ the total gas mass,
$M_g^\mathrm{IN}$
the gas mass within $r_t$, 
$M_g^\mathrm{OUT}$ the mass of the (stripped) gas outside $r_t$, we have

\begin{equation}
\frac{M_g^\mathrm{IN}}{M_g}=
\frac{ \int_0^{r_t}r\Sigma_g \mathrm{d}r }
{\int_0^{\infty}r\Sigma_g \mathrm{d}r } =
1-\left[ 
e^{-r_t/R_{d,g}} \left( \frac{r_t}{R_{d,g}} +1 \right)
\right]
\end{equation}

\begin{equation}
\frac{M_g^\mathrm{IN}}{M_g}=
1-\frac{M_g^\mathrm{OUT}}{M_g}
\end{equation}
therefore, if we call $f_M$ the mass fraction of stripped gas --relative to the total mass-- we have:
\begin{equation}\label{eq:mout}
f_M=\frac{M_g^\mathrm{OUT}}{M_g}=
e^{-r_t/(1.7 R_d)} \left( \frac{r_t}{1.7 R_d} +1 \right)
\end{equation}

Using 
$r_t/R_d$ from Eq.~\ref{eq:rtrd}
in Eq.~\ref{eq:mout},
we have 
obtained an expression for 
the fraction of the stripped gas mass  as a function of
(i) the ICM density $\rho$, 
(ii) the galaxy gas mass fraction $f_{gas}$,
(iii) the galaxy disc scale-length $R_d$, 
(iv) the velocity of the galaxy in the ICM $v$, and
(v) the galaxy stellar mass $M_\star$.
The first one of the above quantities can be expressed as a function
of the cluster velocity dispersion $\sigmacl$ and the clustercentric distance of the galaxy 
in units of $R_{200}$, 
while (ii) and (iii) can be derived with some approximation from the 
galaxy stellar mass $M_\star$ under the assumptions described below.

The ICM density $\rho$ is calculated assuming a $\beta$ model:
\begin{equation}\label{eq:rho}
\rho=\rho_0 \left[ 1+\left(\frac{r_\mathrm{cl}}{R_c}\right)^2 \right]^{-3\beta/2}
\end{equation}
where $\rho_0$ is the gas density at the centre of the cluster, $R_c$ is the cluster core radius,
and $r_\mathrm{cl}$ is the distance of the galaxy from the cluster centre.
We linearly interpolated the 
values in Table 1 from \cite{gaspIX}, taking into account the model revision
described in \cite{gaspIXerr}
to get an expression of
$\rho_0$ and $R_c$ (in units of $R_{200}$) as a function of the cluster velocity dispersion $\sigmacl$\footnote{The choice of this approach was also driven by the fact that $beta$ model parameters are not available for all target clusters}:

\begin{equation}
\rho_0=-3.686\times 10^{-3}\sigmacl+6.200
\end{equation}

\begin{equation}
R_c/R_{200}=6.738\times 10^{-5} \sigmacl-3.157\times10^{-2}
\end{equation}
with $\sigmacl$ in km/s and $\rho_0$ in $10^{-23}$ kg/m$^{3}$.
We assumed $\beta=0.5$ for all clusters.
We therefore defined $\rho(r)$ as a function of $\sigmacl$ and the distance from the cluster centre in units of $R_{200}$.
\begin{equation}
\begin{split}
\rho=&(-3.686\times 10^{-3}\sigmacl+6.200)\; \times\\
&\left[ 1+\left(\frac{r_\mathrm{cl}/R_{200}}{
6.738\times 10^{-5} \sigmacl-3.157\times10^{-2}
}\right)^2 \right]^{-0.75}
\end{split}
\end{equation}

Following \cite{gaspIX}, we can express
$f_\mathrm{gas}$ as a function of the galaxy mass; using the results
of \cite{popp+2014} --obtained considering both \ion{H}{1} and H$_2$-- we adopted the following quadratic relation:
\begin{equation}
f_\mathrm{gas}=
0.158 \, (\log M_\star/M_{\sun})^2-3.548\, \log M_\star / M_{\sun}+  19.964
\end{equation}

We lastly assume the scaling relation between the stellar disc scale-length and the stellar mass
of galaxies from \cite{wu2018}:
\begin{equation} \label{eq:rd}
\log R_d= 0.321 \times \log \left( M_\star/M_{\sun} \right) +0.343
\end{equation}
with $R_d$ in kpc.

\begin{figure}
\centering
\includegraphics[width=\columnwidth]{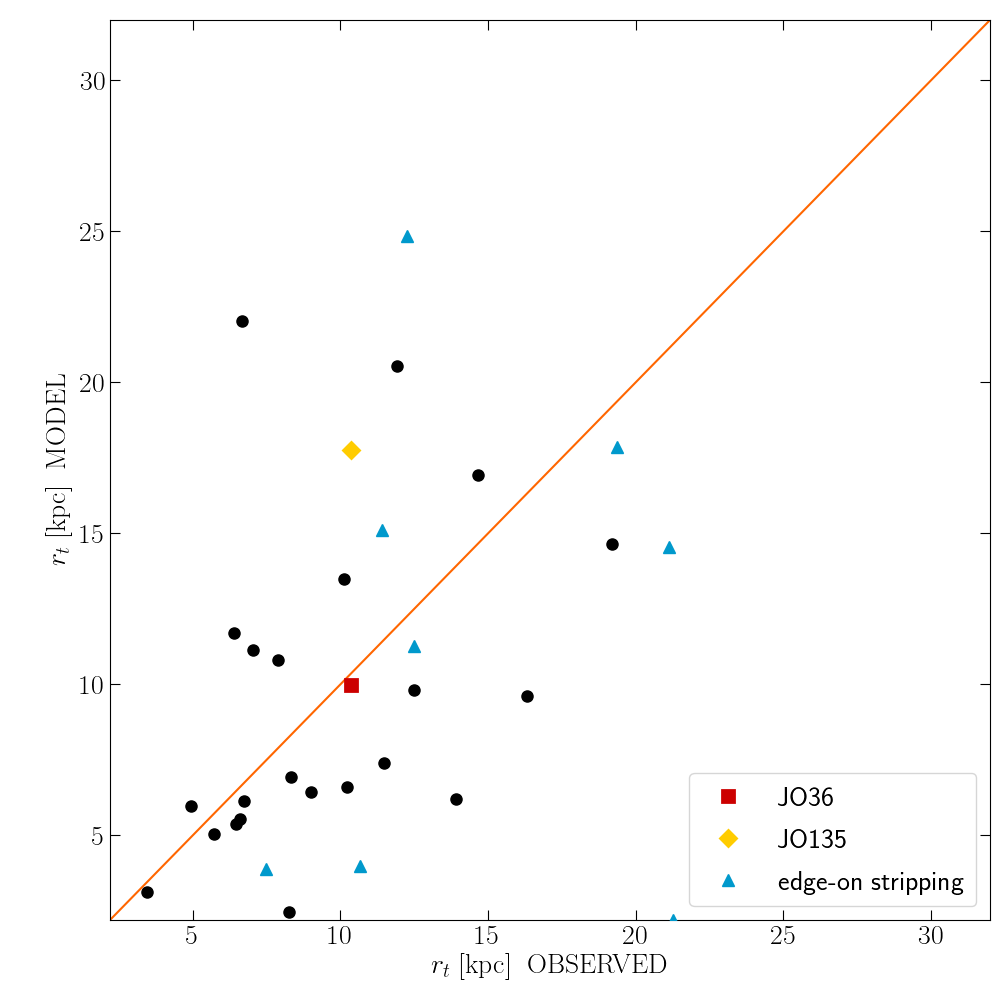}\\
\caption{The truncation radius $r_t$ computed using the model described in Sect. \ref{sec:model} compared with the value estimated from our observations.
The diagonal line is the 1:1 relation.
Symbols are as in Fig.\ref{fig:model_obs}}
\label{fig:rt_all}
\end{figure}

As a sanity check,
in Fig.~\ref{fig:rt_all} we compare the $r_t$ computed by using 
Eq.~\ref{eq:rtrd} (with $R_d$ from Eq.~\ref{eq:rd}) with the value estimated from our MUSE data; this was defined as the maximum extension of the H$\alpha$ emission along the galaxy major axis.
The overall agreement between the observed $r_t$ and the value obtained by our model is satisfactory and  supports the reliability of our modeling.
The scatter in Fig. \ref{fig:rt_all}
originates both from projection effects in the derivation of the observed stripping radius, as well as the caveats inherent to the simple modelization.
Figure \ref{fig:rt} in the Appendix shows examples of the the computed $r_t$ compared with the H$\alpha$
distribution for galaxies with extended tails.

\begin{figure}
\centering
\includegraphics[width=\columnwidth]{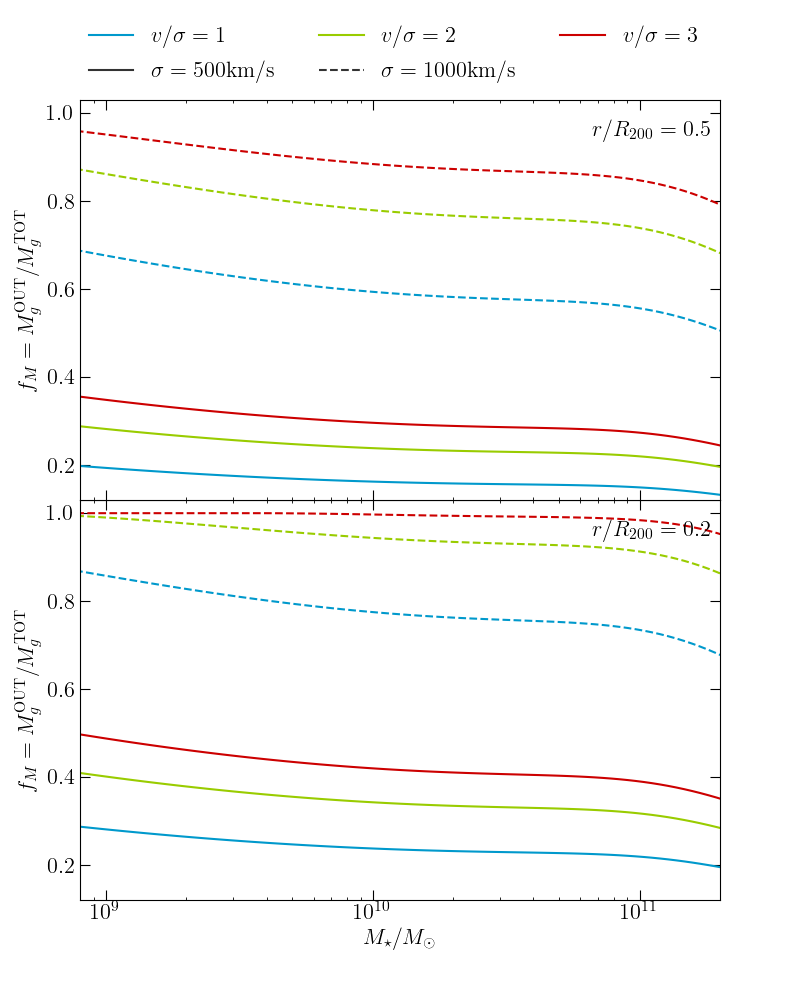}\\
\caption{The two panels show the computed fraction of the gas mass stripped by ram-pressure as a function of the stellar mass for galaxies. As shown on the legend at the top of the figure, solid and dashed lines show the case for galaxies infalling into clusters of different velocity dispersion ($\sigmacl$) while the line colours indicate the value of the galaxy velocity (in units of $\sigma$) in the ICM. In the lower and in the upper panel we plot the values obtained for galaxies located respectively at 0.2 and 0.5 $R_{200}$ from the host cluster centre.}
\label{fig:models}
\end{figure}

To conclude, using Eq.~\ref{eq:rtrd} and Eq.~\ref{eq:mout}, and the above mentioned assumptions, we can compute the mass fraction of stripped gas $f_M$ as a function of cluster velocity dispersion $\sigmacl$, galaxy peculiar velocity $v$, clustercentric distance $r_{\mathrm{cl}}/R_{200}$, and stellar mass $M_\star$. 
A general view of the results obtained from the analytical model
is presented in Fig.~\ref{fig:models}. At fixed values of all other parameters,
the fraction of the stripped gas mass $f_M$ decreases for galaxies of increasing stellar mass; the slope of each curve is however rather shallow, indicating that the galaxy mass is not a driving parameter for the fraction of stripped gas. The increase of $f_M$ at increasing galaxy speed and host cluster mass due to the stronger ram-pressure is shown by the different sets of lines; the two panels show two different test cases for galaxies located at different clustercentric distances, to highlight the effect of the increased ram-pressure in the inner regions of the clusters on $f_M$.

z

We now want to estimate the fraction of star formation which is in the tail from the gas stripped fraction. This will depend on the star formation efficiency, i.e. the amount of stars formed per unit of gas mass, and whether this varies from the disk to the tails.
If the star formation efficiency
were constant throughout the galaxy
(in particular if there were no variations between the disc and the tails), then the fraction of SFR in the tail $\fo$  would be equal to the mass fraction of the stripped gas $f_M$:
\begin{equation}\label{eq:model_obs}
f_M=
\kappa \fo =
\kappa \frac{\sfro}{\sfrt}
\end{equation}
with $\kappa=1$.

A constant $\kappa = 1$ is a useful simplification, although observations find that the SFE decreases as a function of galaxy radius (e.g. \citealt{yim+2014}), so would be lower in the outer regions where gas is more likely to be removed.  Indeed, molecular gas observations of stripped tails have suggested that the star formation efficiency is lower in the tails than in the disks 
\citep{jach+2014,jach+2017,verd+2015,gaspX,more+2020}.
 Using Fig.~6 from \citet{gaspX}, based on CO(2-1) APEX observations of four GASP galaxies, 
 we estimate that the difference in star formation efficiency is a factor of $\sim4$. 
 This is also confirmed by a recent analysis of ALMA data \citep{more+2020ALMA}.
In addition, by combining the results of our APEX and ALMA observations with \ion{H}{1} measurements from JVLA of JO206 \citep{gaspXVII} we find that the total star formation efficiency (star formation per unit of molecular+neutral gas mass) is lower in the tail than in the disk by a factor 5.4.
 
\begin{figure}
\centering
\includegraphics[width=\columnwidth]{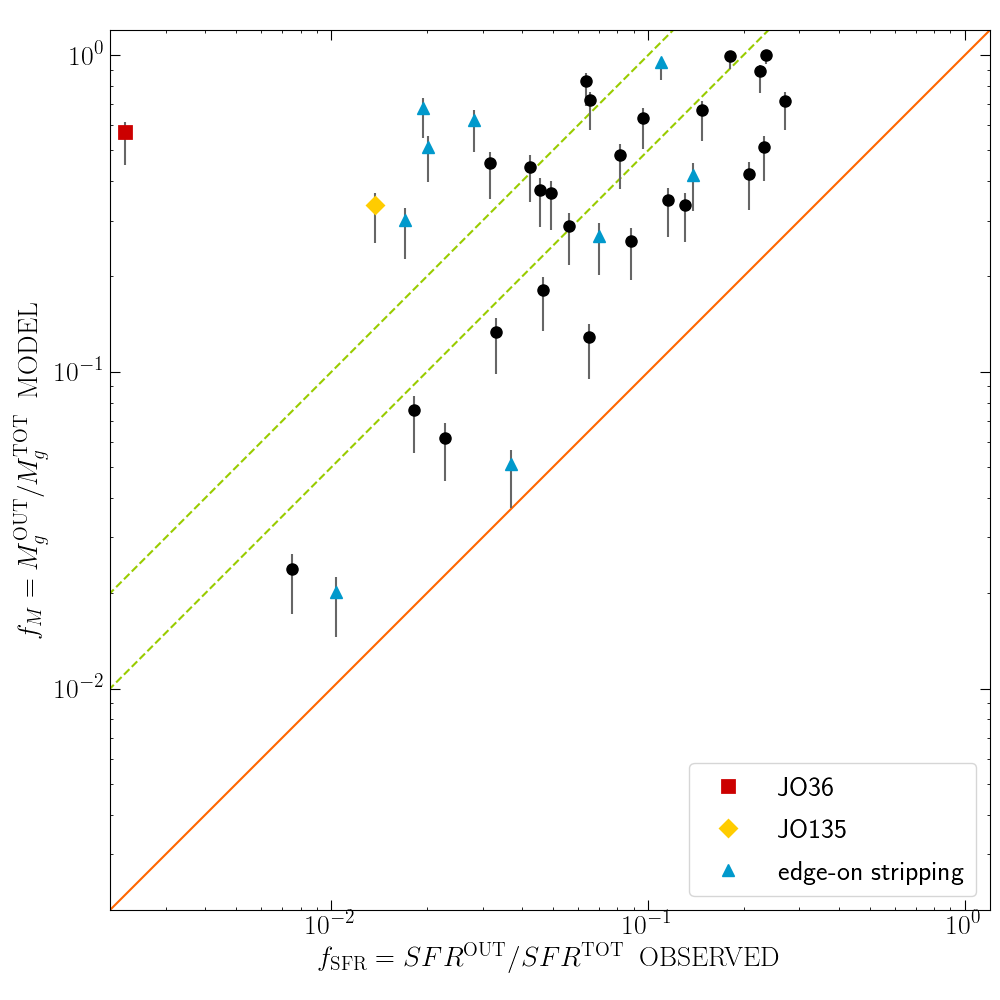}\\
\caption{The  stripped mass fraction computed with our model is plotted against the  observed values of the SFR fraction in the tails of stripped galaxies (see text). 
Galaxies that are likely undergoing edge-on stripping are shown with triangles.
Two galaxies, namely JO36 and JO135 are shown with different symbols and are discussed separately in the text.
The solid line, shown for comparison, corresponds
to $\kappa=1$ in Eq.~\ref{eq:model_obs} (the 1:1 relation), while the two dashed lines correspond to $\kappa=$5 and 10.}
\label{fig:model_obs}
\end{figure}

 In Fig.~\ref{fig:model_obs} we compare the $f_M$
computed with our model with the observed $\fo$ values for galaxies with 
$\sfro>10^{-2}\msunyr$. The data-points are clearly not distributed on the $f_M=\fo$ relation, and the mean value of $\kappa$ (see Eq.~\ref{eq:model_obs}) that we obtain from our data --excluding post-stripping and edge-on stripping galaxies-- is 5.3 (median value 4.5), in striking agreement with the difference in star formation efficiency between tails and disks.

Thus, we observe a correlation between the observed SFR fraction in the tails and the expected mass fraction of the stripped gas based on our model; the relation between these two is compatible with the star formation efficiency measured from the GASP gas studies.

The scatter around this relation is large, and
in the following we discuss some caveats that must be taken into consideration to properly interpret this result. \begin{itemize}
    \item
    As already noted in \cite{gaspIX},  our approach likely overestimates RPS:  galaxy models assume a pure disc profile and this may under-estimate the  anchoring force by neglecting the contribution of the dark matter halo and bulge. The fraction of the stripped gas mass would then be over-estimated, shifting upwards the data-points in Fig.~\ref{fig:model_obs}.
    \item 
    Our analytical results refer to the case of galaxies falling nearly face-on into the ICM;  in other cases, the ram-pressure would be lower than what is obtained assuming  \cite{gunn+1972} prescriptions. Consequently, our model is expected to overestimate the mass of stripped gas in the case of edge-on stripping. Not all edge-on stripped galaxies (blue triangles), however, lie above the median in Figure~8.
    \item
    A fraction of the stripped gas may be completely mixed with the ICM and/or a fraction of the ionised gas emission may be below the MUSE detection limit. Therefore, the gas in the stripped tail at the moment we observe it may be just a fraction of the gas ever stripped
    and the computed $f_M$ would over-estimate the
    observed $\fo$. This effect is clearly more important for  galaxies in a late stage of stripping, or even more for post-stripping galaxies. JO36 has a small tail and a truncated H$\alpha$ disc (see \citealt{gaspIII} for a detailed study of JO36); for this galaxy (red square in \ref{fig:model_obs}) in fact we measured a very low fraction of SFR in the tail ($<$1\%) while our model predicts a consistent fraction of stripped gas. The most plausible scenario is therefore that most of the gas stripped from the galaxy is already lost and dispersed in the ICM and just a very minor fraction of it is close to the galaxy and in dense regions still able to form stars. 
    The other 3 GASP galaxies with truncated H$\alpha$ disc (JO10, JO23, and JW108) are not included in Fig.~\ref{fig:model_obs} because they have 
    $\sfro<10^{-2}\msunyr$
    and they would be placed in Fig.~\ref{fig:model_obs} even more to the left than JO36.
    
    In Fig.~\ref{fig:model_obs} we also note another outlier that lies in the upper-left region of the diagram. It is JO135 (shown as a yellow diamond). This massive galaxy has a rather long tail of ionised gas but 
    we measure a low SFR in the tail ($\sfro=0.03 \msunyr$, $\fo\sim1\%$).
    This is likely
    due to the fact that part of the gas in its tail is
    ionised by the radiation from the central AGN \citep{gaspVI,gaspXIX} and this was consequently not considered in the computation of the total SFR in the tail. The measured $\sfro$ could therefore be substantially under-estimated.
    \item
    Our results are affected by projection effects; both the measured radial velocity and the clustercentric distance component underestimate the 3D values. This induces opposite effects on the computed ram-pressure (an under-estimated speed implies an under-estimated ram-pressure while an under-estimated distance implies an over-estimated ram-pressure). To evaluate the impact of projection effects on our results and on the dispersion of the data-points in Fig.~\ref{fig:model_obs}, we re-computed $f_M$ for each galaxy assuming a velocity $2\times$ the measured radial velocity and then assuming a clustercentric distance $2\times$ the measured component on the plane of the sky. The two resulting values for each galaxy are shown by the vertical bars in Fig.~\ref{fig:model_obs}.
    We conclude that overall the distribution of the data-points is not significantly affected by projection effects.
\end{itemize}

Finally, we emphasise that our model is based on many approximations and strong assumptions and was developed to provide a description of general trends.
We assumed a general relation to describe the gas distribution without any assumption on the  
spatial distribution of the different gas phases. In general, observations of undisturbed galaxies show that molecular gas dominates in the inner regions and atomic gas in the outer ones \citep[see e.g.][]{bigi+2012}.
Our assumptions are compatible with the total gas scale length found by \cite{bigi+2012} (see footnote 1);
further investigation of this will be carried out using data from
our ongoing multi-wavelength observing campaign. This will probe atomic and molecular gas in GASP galaxies with a resolution similar to the one we obtained for ionised gas
with MUSE and will allow us to investigate in detail the spatially resolved
SFE.
Fig.~\ref{fig:model_obs} shows that our approach provides a quite satisfactory description of the observations, which in turn implies that, albeit with a large scatter, the four quantities that can be derived from observations (cluster velocity dispersion, galaxy velocity, clustercentric distance and mass) can provide a crude approximation of the fraction of star formation taking place in the tails. This also suggests that additional factors (e.g. link with cluster substructure/merging) 
are probably only second order effects.

\section{intracluster light}\label{sec:icl}

In this section, the GASP results will be used to obtain a rough estimate of the total  contribution from RPS to the intracluster light. An implicit assumption
we will make is that the stars formed in the regions which we call tails will be lost from the galaxy at some stage and will become part of the intracluster component, thus neglecting the fact that some of these stars might still be bound to the galaxy and eventually join the disk. Our generous choice of the disk boundaries should limit this effect, but this caveat should be kept in mind.

We used the complete catalogue of candidate RPS galaxies published from \citet{pogg+2016}, from which the GASP target galaxies were selected. This catalogue was compiled using data from WINGS \citep{fasa+2006} and OmegaWINGS \citep{gull+2015}, which are two complete surveys of X-ray selected
clusters in the redshift range 0.04-0.07 at galactic latitude 
$ \vert l\vert>20^{\circ}$ (66\% of the sky). 
The selection of GASP candidates was carried out trying to span the whole range of parameters of interest, in particular
galaxy mass, cluster mass and JClass \citep[degree of asymmetry in the optical galaxy morphology, see][]{pogg+2016}.
We can therefore assume that GASP provides a reasonably representative snapshot of the population of ram-pressure stripped galaxies in the nearby Universe.

\begin{deluxetable}{rrrrr}
\tablecaption{
Integrated values of the SFR for all galaxies grouped according to
the JClass classification from \citet{pogg+2016}.
Columns are:
1) JClass;
2) total number of galaxies in \citet{pogg+2016};
3) number of galaxies in this paper; 
4) average of the $\sfro$ in the tail for galaxies in this paper;
5) total $\sfro$ multiplied by the number of galaxies in \citet{pogg+2016}.
\label{tab:sfrJcl}
}
\tablehead{
\colhead{JClass}& 
\colhead{$n^{\mathrm{P16}}_{\mathrm{JC}}$}&
\colhead{$n_{\mathrm{JC}}$}&
\colhead{$\sfro_{\mathrm{JC}}$}& 
\colhead{$\sfro_{\mathrm{JC,TOT}}$}\\
&&&
\colhead{$M_{\sun}\mathrm{yr}^{-1}$}&
\colhead{$M_{\sun}\mathrm{yr}^{-1}$}
}
\startdata
     1 & 131&  13 &0.031 &  4.11\\
     2 & 115&  10 &0.012 &  1.41\\
     3 &  67&  11 &0.035 &  2.37\\
     4 &  21&  12 &0.089 &  1.88\\
     5 &  10&   8 &0.595 &  5.95\\
\enddata
\end{deluxetable}

To estimate the total amount of SFR in the tails of ram-pressure stripped galaxies we grouped the GASP galaxies according to the JClass;
for each of the five groups we computed the average
tail SFR6 that we call $\sfro_{\mathrm{JC}}$. 
We then multiplied the resulting values
for the number of galaxies in
each JClass in \citet{pogg+2016}. 
Results are reported in Table~\ref{tab:sfrJcl}.
The total SFR for all ram-pressure stripped galaxies is the sum
of the five values obtained for each JClass:

\begin{equation}
\sfro_{\mathrm{TOT}}=
\sum_{\mathrm{JC}=1}^5 
n^{\mathrm{P16}}_{\mathrm{JC}} \times
\sfro_{\mathrm{JC}}=
\sum_{\mathrm{JC}=1}^5 
\sfro_{\mathrm{JC,TOT}}
\end{equation}

The resulting estimate for the integrated SFR in the tails of all ram-pressure stripped galaxies in these clusters is
$15.72 M_{\sun} \mathrm{yr}^{-1}$. Since the total number of clusters hosting ram-pressure stripping candidates in \cite{pogg+2016} is 71, the average value per cluster is $0.22 M_{\sun} \mathrm{yr}^{-1}$.

We note that the sky coverage of our target clusters in not uniform, as the 
FoV of WINGS and OmegaWINGS imaging is $30^\prime$ and $1^\circ$, respectively, and only a fraction of clusters has OmegaWINGS observations. To assess the possible impact of this on our analysis, we repeated our computation by selecting GASP and \citet{pogg+2016} data only for galaxies in clusters with OmegaWINGS observations. In this case we found an average SFR per cluster of $0.25 M_{\sun} \mathrm{yr}^{-1}$. This value is not significantly different from the one we obtain from our complete dataset, showing that the different sky coverage of the WINGS/OmegaWINGS target clusters does not affect our conclusions.

 We can now use our GASP results on the SFR in the tails to trace back in time the contribution of ram-pressure stripping to the ICL.
This can be computed assuming that the average value of the SFR in the tail per cluster is simply proportional to the infalling rate of 
galaxies in the cluster. 
We consider the infall of galaxies at $z \leq 1$, an epoch at which the evolution of the ICM is negligible and WINGS/OMEGAWINGS clusters should have already developed their dense and hot ICM able to induce RPS \citep{leauthaud+2010, bulbul+2019}.

\begin{figure}
\centering
\includegraphics[width=\columnwidth]{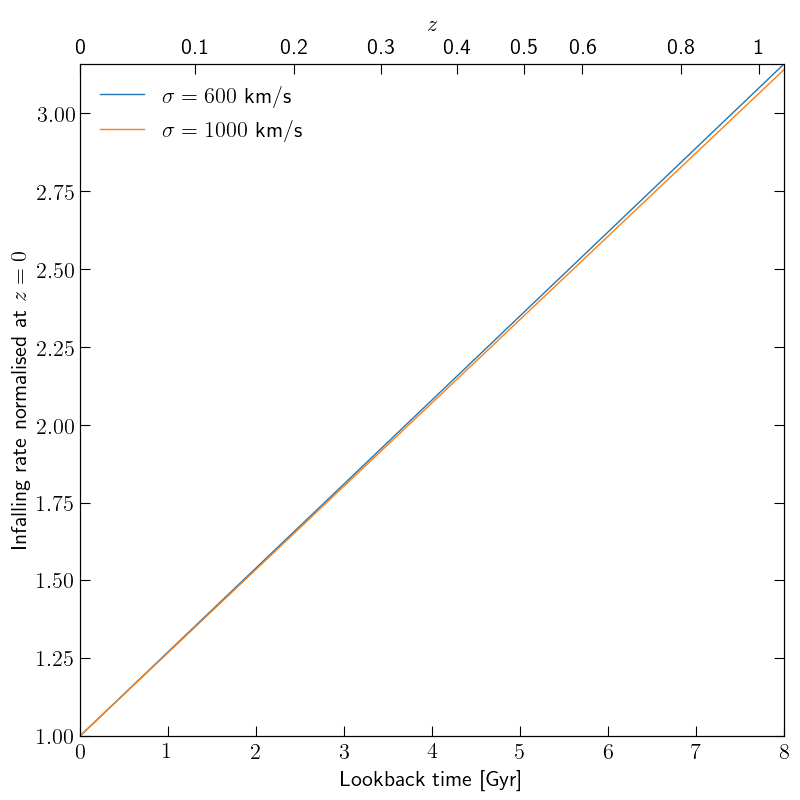}\\
\caption{
Infalling rates normalised to the value at $z=0$ for the two model clusters.
}
\label{fig:infallingrates}
\end{figure}

We compute the rate of galaxies infalling into galaxy clusters as a function of lookback time by using the semi-analytic model of \citet{Henriques+2015}. This model is based on implementing analytic equations which simplify the baryonic physics of galaxy formation on the background of a cosmological Millennium N-body simulation \citep{Springel+2005} which has been recalibrated to be consistent with the cosmological model favoured by the Planck satellite \citep{Planck_XVI+2014}. This model assumes a $\sigma_8$ = 0.829, and has cosmological parameters closer to those used in our observations than the original Millennium simulation. The small differences in our adopted $\Omega_M$ and $\Omega_\Lambda$ do not have appreciable differences in the infall histories recovered. 

This semi-analytic model is the most recent version of the Munich galaxy formation model, and was particularly focused on tweaking the analytics implementations to correctly reproduce star formation rates, colours and stellar masses of galaxies. While semi-analytic models can vary in their underlying equations, and therefore in their predictions, in this work we only use the predictions for the stellar masses of galaxies at any particular redshift which are most robustly predicted by different models. Indeed, we recover similar results 
when using a recent version of the Durham galaxy formation model for the same cosmology.

To take into account the possible influence of the cluster halo mass on the infalling rate, we calculated it for two different halo masses: one corresponding to a low-mass cluster ($\sim 3 \times 10^{14} M_\sun$) and one to high-mass one ($\sim 2 \times 10^{15} M_\sun$). 
Therefore, from the Henriques et al. models we select all halos having a velocity dispersion between 500 and 600 km/s and 800 and 1200 km/s, respectively.  We then selected, within each halo, every galaxy with a stellar mass M$_\mathrm{stellar} >$ M$_\mathrm{stellar,cut}$ and track it through the simulation to find the time at which it was first accreted into the final cluster. The results were examined for M$_\mathrm{stellar,cut}$ = 10$^{10}$ and 10$^9$  and in both total halo mass bins. 
To exclude from our analysis galaxies that were pre-processed before being accreted into the halo, we considered for our calculations only objects at first infall. This selection was accomplished by ignoring galaxies which were in a dark matter halo with a velocity dispersion of $<$ 500 km/s at the time of their accretion into the main cluster.

The cumulative fractions of galaxies infalling into the two simulated clusters as a function of lookback time were fitted using a second-order polynomial.
The infalling rates are obtained as the derivative of the cumulative fraction of infalling galaxies and therefore result to be a linear function of the look-back time.
As expected the infalling rates are larger for the high-mass halo. 
However, we are only interested in the evolution of the infall rate, not its absolute value, so we can normalize it to the infalling rate at z=0, which roughly represents the epoch of the GASP observations.
After this normalization, the evolution of the infall rate is almost independent of the cluster velocity dispersion bin or the stellar mass cut. As this normalised infalling rate is essentially driven by the underlying cosmology, the mass independence is not surprising.  
The results are shown in Fig.~\ref{fig:infallingrates}.

The computed normalized infalling rates, simply multiplied by our GASP estimate of the average total SFR in the tails per cluster ($0.22 M_{\sun} \mathrm{yr}^{-1}$, see above) gives our estimate of the evolution of the contribution of ram-pressure stripping to the ICM per average cluster. At $z=1$ it results to be $\approx$3 times the value at $z=0$. By integrating our results we estimate  a total value of $\approx 4 \times 10^9 M_{\sun}$ of stars formed per cluster in the ICM from ram-pressure stripped gas since z$\sim 1$. We stress that this value has been derived for a set of clusters with an average value of velocity dispersion of $750 \kms$ at $z \sim 0$ \citep{pogg+2016}, while this number could be higher for more massive clusters.

The contribution of ram-pressure stripping in shaping the ICL is still extremely uncertain. \cite{adam+2016} concluded that RPS is the most plausible process generating the ICL sources;
other studies proposed different mechanisms, such as tidal stripping of massive  \citep{dema+2018, mont+2018} or low-mass \citep{mori+2017} galaxies.  Direct measurements of the ICL are challenging, mostly because it is extremely difficult to disentangle the diffuse component of the ICL from the contribution of the galaxies and in particular from the Brightest Cluster Galaxy (BCG). A clear understanding of the origin of the ICL is also hampered by the uncertainties on the properties of the mass, the age and metallicity of the ICL's stellar component. Most of the literature studies of the ICL are in fact based on broad-band imaging data and photometric SED fitting. Only a few spectroscopic studies have been published so far and it is not possible to draw final conclusions on the properties and the origin of the ICL (see e.g. \citealt{cocc+2011, meln+2012} based on long-slit/MOS spectroscopy and \citealt{adam+2016} based on MUSE IFU spectroscopy). More spectroscopic observations are still required to further investigate the physical processes driving the formation and the evolution of the ICL. The exceptional spatial resolution, FoV and sensitivity of MUSE could most likely play a primary role.

\section{Summary and conclusions}\label{sec:conclusions}
As part of the GASP project based on an ESO Large Programme with MUSE, this paper focuses on the SFR in the tails of cluster galaxies underging ram-pressure stripping. By considering all GASP cluster galaxies --excluding merging and tidally interacting systems -- we used a sample of 54 galaxies; our sample covers a wide parameter space in terms of galaxy stellar mass, between less than $10^9$ and $10^{11.5} M_\sun$, and host cluster mass/velocity dispersion, between 400 to more than 1000 $\kms$. We defined a method to conservatively define a mask to disentangle the ram-pressure stripped gas tail from the galaxy main body. We computed the SFR from the H$\alpha$ emission by using BPT diagnostic diagrams to exclude the gas not ionised by SF. We used our measurements to study how the SFR in the tail depends on the properties of the galaxy and of its host cluster.  We found that there is not a single dominant parameter driving the observed value of the $\sfro$; the mass of the galaxy, its position and velocity in the host cluster and all the parameters defining the distribution of the ICM density are all to be considered to properly account for the $\sfro$. However we found general trends that are here summarized.

 All galaxies with marginal $\sfro$ are found at relatively large distances from the host cluster center. Some of these are galaxies at the first infall in the cluster that are being accreted and therefore had not been stripped yet.
    
All the galaxies with large $\sfro$ ($>0.25\msunyr$) that are moving  at large speed in the innermost regions of the clusters are massive and hosted in low-mass clusters; only under these conditions the gravitational potential can contrast the extreme ram-pressure stripping that would otherwise strip most of the gas before these galaxies could reach the inner cluster regions

RPS occurs preferentially at intermediate clustercentric distance in massive clusters and at lower distances in intermediate and low-mass clusters. This is because galaxies are nearly completely stripped when they reach the dense region of high-mass clusters.

To provide a method to predict the amount of SFR in the tails using observable quantities, based on our observational results. We developed a simple analytical approach based on ram-pressure prescriptions from \cite{gunn+1972}; we aimed at deriving the fraction of stripped mass as a function of galaxy and clusters parameters that can be easily obtained from observations. Following other literature work \cite[e.g.][]{smit+2012,ower+2019,gaspIX}, 
we made standard assumptions and
we adopted scaling relations for (i) the galaxy gas fraction and disc scale-length as a function of the galactic stellar mass and (ii) the ICM central density and core radius as a function of the cluster velocity dispersion. As a result, we obtained an analytic expression for the mass fraction of stripped gas as a function of four parameters: 
the cluster velocity dispersion, the galaxy stellar mass, its clustercentric distance and speed in the ICM. To assess the reliability of our model, we compared the computed truncation radius --the galactocentric distance at which the ram-pressure equals the gravitational anchoring force-- with the ionised gas emission maps obtained from MUSE observations; the remarkable agreement is a strong indication that our assumptions provides a reasonable description of the properties of the galaxies and their host cluster.

A direct comparison of the fraction of stripped mass computed with our model with the observed fraction of SFR in the tails shows a very good agreement (albeit with a large scatter) between the two quantities if the total (molecular+neutral) star-formation efficiency is lower in the tail than in the disk by a factor $\sim 5$, in excellent agreement with the efficiency derived from our ongoing CO and \ion{H}{1} observing campaign  with APEX, ALMA and JVLA \citep{gaspX,gaspXVII,more+2020ALMA}.

We used the values of the SFR in the tails of stripped gas to  estimate the contribution of RPS to the ICL. By statistically correcting our GASP measurements using the whole GASP parent candidate catalog from \cite{pogg+2016}, we found that the average SFR for all ram-pressure stripped galaxies per cluster is $0.22\msunyr$. We finally used this result to extrapolate the contribution to the ICL at different look-back times, by assuming that it is proportional to the number of galaxies at first infall into the cluster.
The infalling rate was computed using the cosmological semi-analytical model by \cite{Henriques+2015} based on the Millenium N-body Simulation.
We estimated a total average value per cluster of $\approx 4\times 10^9M_\sun$ of stars formed in the ICM from ram-pressure stripped gas since $z=1$. This estimate can be used to evaluate the contribution of RPS in shaping the ICL and therefore is a valuable contribution to the still open debate about the physical processes driving the formation and the evolution of the ICL.

\acknowledgments
We would like to warmly thank Andrea Biviano for the useful discussions during the preparation of the manuscript.
Based on observations collected at the European Organisation for Astronomical Research in the Southern
Hemisphere under ESO programme 196.A-0578. This
work made use of the KUBEVIZ software which is publicly
available at \url{http://www.mpe.mpg.de/~dwilman/kubeviz}.
This research made use of Astropy, a community-developed core Python package for Astronomy (Astropy Collaboration, \citeyear{astropy}).
This project has received funding from the European Research Council (ERC) under the European Union's Horizon 2020 research and innovation programme (grant agreement No. 833824).
We acknowledge financial support from PRIN-SKA 2017 (PI L. Hunt)
and "INAF main-streams" funding programme (PI B. Vulcani).
Y.J. acknowledges financial support from CONICYT PAI (Concurso Nacional de Inserci\'on en la Academia 2017) No. 79170132 and FONDECYT Iniciaci\'on 2018 No. 11180558.
M.G., B.V. and D.B. acknowledge the support from grant PRIN MIUR 2017 - 20173ML3WW\_001
\bibliographystyle{aasjournal}
\bibliography{bib}

\begin{thebibliography}{}
\expandafter\ifx\csname natexlab\endcsname\relax\def\natexlab#1{#1}\fi
\providecommand{\url}[1]{\href{#1}{#1}}

\bibitem[{{Abramson} {et~al.}(2011){Abramson}, {Kenney}, {Crowl}, {Chung}, {van
  Gorkom}, {Vollmer}, \& {Schiminovich}}]{abra+2011}
{Abramson}, A., {Kenney}, J. D.~P., {Crowl}, H.~H., {et~al.} 2011, \aj, 141,
  164

\bibitem[{{Adami} {et~al.}(2016){Adami}, {Pompei}, {Sadibekova}, {Clerc},
  {Iovino}, {McGee}, {Guennou}, {Birkinshaw}, {Horellou}, {Maurogordato},
  {Pacaud}, {Pierre}, {Poggianti}, \& {Willis}}]{adam+2016}
{Adami}, C., {Pompei}, E., {Sadibekova}, T., {et~al.} 2016, \aap, 592, A7

\bibitem[{{Bacon} {et~al.}(2010){Bacon}, {Accardo}, {Adjali}, {Anwand},
  {Bauer}, {Biswas}, {Blaizot}, {Boudon}, {Brau-Nogue}, {Brinchmann},
  {Caillier}, {Capoani}, {Carollo}, {Contini}, {Couderc}, {Daguis{\'e}},
  {Deiries}, {Delabre}, {Dreizler}, {Dubois}, {Dupieux}, {Dupuy}, {Emsellem},
  {Fechner}, {Fleischmann}, {Fran{\c c}ois}, {Gallou}, {Gharsa}, {Glindemann},
  {Gojak}, {Guiderdoni}, {Hansali}, {Hahn}, {Jarno}, {Kelz}, {Koehler},
  {Kosmalski}, {Laurent}, {Le Floch}, {Lilly}, {Lizon}, {Loupias}, {Manescau},
  {Monstein}, {Nicklas}, {Olaya}, {Pares}, {Pasquini}, {P{\'e}contal-Rousset},
  {Pell{\'o}}, {Petit}, {Popow}, {Reiss}, {Remillieux}, {Renault}, {Roth},
  {Rupprecht}, {Serre}, {Schaye}, {Soucail}, {Steinmetz}, {Streicher}, {Stuik},
  {Valentin}, {Vernet}, {Weilbacher}, {Wisotzki}, \& {Yerle}}]{baco+2010}
{Bacon}, R., {Accardo}, M., {Adjali}, L., {et~al.} 2010, in \procspie, Vol.
  7735, Ground-based and Airborne Instrumentation for Astronomy III, 773508

\bibitem[{{Baldwin} {et~al.}(1981){Baldwin}, {Phillips}, \& {Terlevich}}]{bpt}
{Baldwin}, J.~A., {Phillips}, M.~M., \& {Terlevich}, R. 1981, \pasp, 93, 5

\bibitem[{{Bellhouse} {et~al.}(2017){Bellhouse}, {Jaff{\'e}}, {Hau}, {McGee},
  {Poggianti}, {Moretti}, {Gullieuszik}, {Bettoni}, {Fasano}, {D'Onofrio},
  {Fritz}, {Omizzolo}, {Sheen}, \& {Vulcani}}]{gaspII}
{Bellhouse}, C., {Jaff{\'e}}, Y.~L., {Hau}, G.~K.~T., {et~al.} 2017, \apj, 844,
  49

\bibitem[{{Bellhouse} {et~al.}(2019){Bellhouse}, {Jaff{\'e}}, {McGee},
  {Poggianti}, {Smith}, {Tonnesen}, {Fritz}, {Hau}, {Gullieuszik}, {Vulcani},
  {Fasano}, {Moretti}, {George}, {Bettoni}, {D'Onofrio}, {Omizzolo}, \&
  {Sheen}}]{gaspXV}
{Bellhouse}, C., {Jaff{\'e}}, Y.~L., {McGee}, S.~L., {et~al.} 2019, \mnras,
  485, 1157

\bibitem[{{Bigiel} \& {Blitz}(2012)}]{bigi+2012}
{Bigiel}, F., \& {Blitz}, L. 2012, \apj, 756, 183

\bibitem[{{Biviano} {et~al.}(2017){Biviano}, {Moretti}, {Paccagnella},
  {Poggianti}, {Bettoni}, {Gullieuszik}, {Vulcani}, {Fasano}, {D'Onofrio},
  {Fritz}, \& {Cava}}]{bivi+2017}
{Biviano}, A., {Moretti}, A., {Paccagnella}, A., {et~al.} 2017, \aap, 607, A81

\bibitem[{{Boissier} {et~al.}(2012){Boissier}, {Boselli}, {Duc}, {Cortese},
  {van Driel}, {Heinis}, {Voyer}, {Cucciati}, {Ferrarese}, {C{\^o}t{\'e}},
  {Cuillandre}, {Gwyn}, \& {Mei}}]{bois+2012}
{Boissier}, S., {Boselli}, A., {Duc}, P.~A., {et~al.} 2012, \aap, 545, A142

\bibitem[{{Boselli} \& {Gavazzi}(2006)}]{bose+2006}
{Boselli}, A., \& {Gavazzi}, G. 2006, \pasp, 118, 517

\bibitem[{{Boselli} {et~al.}(2016){Boselli}, {Cuillandre}, {Fossati},
  {Boissier}, {Bomans}, {Consolandi}, {Anselmi}, {Cortese}, {C{\^o}t{\'e}},
  {Durrell}, {Ferrarese}, {Fumagalli}, {Gavazzi}, {Gwyn}, {Hensler}, {Sun}, \&
  {Toloba}}]{bose+2016}
{Boselli}, A., {Cuillandre}, J.~C., {Fossati}, M., {et~al.} 2016, \aap, 587,
  A68

\bibitem[{{Bulbul} {et~al.}(2019){Bulbul}, {Chiu}, {Mohr}, {McDonald},
  {Benson}, {Bautz}, {Bayliss}, {Bleem}, {Brodwin}, {Bocquet}, {Capasso},
  {Dietrich}, \& et~al.}]{bulbul+2019}
{Bulbul}, E., {Chiu}, I.~N., {Mohr}, J.~J., {et~al.} 2019, \apj, 871, 50

\bibitem[{{Calvi} {et~al.}(2011){Calvi}, {Poggianti}, \& {Vulcani}}]{calv+2011}
{Calvi}, R., {Poggianti}, B.~M., \& {Vulcani}, B. 2011, \mnras, 416, 727

\bibitem[{{Cappellari} \& {Emsellem}(2004)}]{capp+2004}
{Cappellari}, M., \& {Emsellem}, E. 2004, \pasp, 116, 138

\bibitem[{{Cardelli} {et~al.}(1989){Cardelli}, {Clayton}, \&
  {Mathis}}]{card+1989}
{Cardelli}, J.~A., {Clayton}, G.~C., \& {Mathis}, J.~S. 1989, \apj, 345, 245

\bibitem[{{Cayatte} {et~al.}(1994){Cayatte}, {Kotanyi}, {Balkowski}, \& {van
  Gorkom}}]{caya+1994}
{Cayatte}, V., {Kotanyi}, C., {Balkowski}, C., \& {van Gorkom}, J.~H. 1994,
  \aj, 107, 1003

\bibitem[{{Chabrier}(2003)}]{chab+2003}
{Chabrier}, G. 2003, \pasp, 115, 763

\bibitem[{{Chung} {et~al.}(2007){Chung}, {van Gorkom}, {Kenney}, \&
  {Vollmer}}]{chun+2007}
{Chung}, A., {van Gorkom}, J.~H., {Kenney}, J. D.~P., \& {Vollmer}, B. 2007,
  \apjl, 659, L115

\bibitem[{{Coccato} {et~al.}(2011){Coccato}, {Gerhard}, {Arnaboldi}, \&
  {Ventimiglia}}]{cocc+2011}
{Coccato}, L., {Gerhard}, O., {Arnaboldi}, M., \& {Ventimiglia}, G. 2011, \aap,
  533, A138

\bibitem[{{Contini} {et~al.}(2018){Contini}, {Yi}, \& {Kang}}]{cont+2018}
{Contini}, E., {Yi}, S.~K., \& {Kang}, X. 2018, \mnras, 479, 932

\bibitem[{{DeMaio} {et~al.}(2018){DeMaio}, {Gonzalez}, {Zabludoff}, {Zaritsky},
  {Connor}, {Donahue}, \& {Mulchaey}}]{dema+2018}
{DeMaio}, T., {Gonzalez}, A.~H., {Zabludoff}, A., {et~al.} 2018, \mnras, 474,
  3009

\bibitem[{{Ebeling} {et~al.}(2014){Ebeling}, {Stephenson}, \&
  {Edge}}]{ebel+2014}
{Ebeling}, H., {Stephenson}, L.~N., \& {Edge}, A.~C. 2014, \apjl, 781, L40

\bibitem[{{Fasano} {et~al.}(2006){Fasano}, {Marmo}, {Varela}, {D'Onofrio},
  {Poggianti}, {Moles}, {Pignatelli}, {Bettoni}, {Kj{\ae}rgaard}, {Rizzi},
  {Couch}, \& {Dressler}}]{fasa+2006}
{Fasano}, G., {Marmo}, C., {Varela}, J., {et~al.} 2006, \aap, 445, 805

\bibitem[{{Fossati} {et~al.}(2016){Fossati}, {Fumagalli}, {Boselli}, {Gavazzi},
  {Sun}, \& {Wilman}}]{foss+2016}
{Fossati}, M., {Fumagalli}, M., {Boselli}, A., {et~al.} 2016, \mnras, 455, 2028

\bibitem[{{Fritz} {et~al.}(2017){Fritz}, {Moretti}, {Gullieuszik}, {Poggianti},
  {Bruzual}, {Vulcani}, {Nicastro}, {Jaff{\'e}}, {Cervantes Sodi}, {Bettoni},
  {Biviano}, {Fasano}, {Charlot}, {Bellhouse}, \& {Hau}}]{gaspIII}
{Fritz}, J., {Moretti}, A., {Gullieuszik}, M., {et~al.} 2017, \apj, 848, 132

\bibitem[{{Fumagalli} {et~al.}(2014){Fumagalli}, {Fossati}, {Hau}, {Gavazzi},
  {Bower}, {Sun}, \& {Boselli}}]{fuma+2014}
{Fumagalli}, M., {Fossati}, M., {Hau}, G.~K.~T., {et~al.} 2014, \mnras, 445,
  4335

\bibitem[{{Gavazzi}(1989)}]{gava1989}
{Gavazzi}, G. 1989, \apj, 346, 59

\bibitem[{{George} {et~al.}(2018){George}, {Poggianti}, {Gullieuszik},
  {Fasano}, {Bellhouse}, {Postma}, {Moretti}, {Jaff{\'e}}, {Vulcani},
  {Bettoni}, {Fritz}, {C{\^o}t{\'e}}, {Ghosh}, {Hutchings}, {Mohan},
  {Sreekumar}, {Stalin}, {Subramaniam}, \& {Tandon}}]{geor+2018}
{George}, K., {Poggianti}, B.~M., {Gullieuszik}, M., {et~al.} 2018, \mnras,
  479, 4126

\bibitem[{{Giallongo} {et~al.}(2014){Giallongo}, {Menci}, {Grazian},
  {Gallozzi}, {Castellano}, {Fiore}, {Fontana}, {Pentericci}, {Boutsia},
  {Paris}, {Speziali}, \& {Testa}}]{gial+2014}
{Giallongo}, E., {Menci}, N., {Grazian}, A., {et~al.} 2014, \apj, 781, 24

\bibitem[{{Giovanelli} \& {Haynes}(1985)}]{giova+1985}
{Giovanelli}, R., \& {Haynes}, M.~P. 1985, \apj, 292, 404

\bibitem[{{Guglielmo} {et~al.}(2015){Guglielmo}, {Poggianti}, {Moretti},
  {Fritz}, {Calvi}, {Vulcani}, {Fasano}, \& {Paccagnella}}]{gugl+2015}
{Guglielmo}, V., {Poggianti}, B.~M., {Moretti}, A., {et~al.} 2015, \mnras, 450,
  2749

\bibitem[{{Gullieuszik} {et~al.}(2015){Gullieuszik}, {Poggianti}, {Fasano},
  {Zaggia}, {Paccagnella}, {Moretti}, {Bettoni}, {D'Onofrio}, {Couch},
  {Vulcani}, {Fritz}, {Omizzolo}, {Baruffolo}, {Schipani}, {Capaccioli}, \&
  {Varela}}]{gull+2015}
{Gullieuszik}, M., {Poggianti}, B., {Fasano}, G., {et~al.} 2015, \aap, 581, A41

\bibitem[{{Gullieuszik} {et~al.}(2017){Gullieuszik}, {Poggianti}, {Moretti},
  {Fritz}, {Jaff{\'e}}, {Hau}, {Bischko}, {Bellhouse}, {Bettoni}, {Fasano},
  {Vulcani}, {D'Onofrio}, \& {Biviano}}]{gaspIV}
{Gullieuszik}, M., {Poggianti}, B.~M., {Moretti}, A., {et~al.} 2017, \apj, 846,
  27

\bibitem[{{Gunn} \& {Gott}(1972)}]{gunn+1972}
{Gunn}, J.~E., \& {Gott}, III, J.~R. 1972, \apj, 176, 1

\bibitem[{{Henriques} {et~al.}(2015){Henriques}, {White}, {Thomas}, {Angulo},
  {Guo}, {Lemson}, {Springel}, \& {Overzier}}]{Henriques+2015}
{Henriques}, B.~M.~B., {White}, S.~D.~M., {Thomas}, P.~A., {et~al.} 2015,
  \mnras, 451, 2663

\bibitem[{{Hester} {et~al.}(2010){Hester}, {Seibert}, {Neill}, {Wyder}, {Gil de
  Paz}, {Madore}, {Martin}, {Schiminovich}, \& {Rich}}]{hest+2010}
{Hester}, J.~A., {Seibert}, M., {Neill}, J.~D., {et~al.} 2010, \apjl, 716, L14

\bibitem[{{J{\'a}chym} {et~al.}(2014){J{\'a}chym}, {Combes}, {Cortese}, {Sun},
  \& {Kenney}}]{jach+2014}
{J{\'a}chym}, P., {Combes}, F., {Cortese}, L., {Sun}, M., \& {Kenney}, J. D.~P.
  2014, \apj, 792, 11

\bibitem[{{J{\'a}chym} {et~al.}(2017){J{\'a}chym}, {Sun}, {Kenney}, {Cortese},
  {Combes}, {Yagi}, {Yoshida}, {Palou{\v{s}}}, \& {Roediger}}]{jach+2017}
{J{\'a}chym}, P., {Sun}, M., {Kenney}, J. D.~P., {et~al.} 2017, \apj, 839, 114

\bibitem[{{J{\'a}chym} {et~al.}(2019){J{\'a}chym}, {Kenney}, {Sun}, {Combes},
  {Cortese}, {Scott}, {Sivanandam}, {Brinks}, {Roediger}, {Palou{\v{s}}}, \&
  {Fumagalli}}]{jach+2019}
{J{\'a}chym}, P., {Kenney}, J. D.~P., {Sun}, M., {et~al.} 2019, \apj, 883, 145

\bibitem[{{Jaff{\'e}} {et~al.}(2015){Jaff{\'e}}, {Smith}, {Candlish},
  {Poggianti}, {Sheen}, \& {Verheijen}}]{jaff+2015}
{Jaff{\'e}}, Y.~L., {Smith}, R., {Candlish}, G.~N., {et~al.} 2015, \mnras, 448,
  1715

\bibitem[{{Jaff{\'e}} {et~al.}(2018){Jaff{\'e}}, {Poggianti}, {Moretti},
  {Gullieuszik}, {Smith}, {Vulcani}, {Fasano}, {Fritz}, {Tonnesen}, {Bettoni},
  {Hau}, {Biviano}, {Bellhouse}, \& {McGee}}]{gaspIX}
{Jaff{\'e}}, Y.~L., {Poggianti}, B.~M., {Moretti}, A., {et~al.} 2018, \mnras,
  476, 4753

\bibitem[{{Jaff{\'e}} {et~al.}(2019){Jaff{\'e}}, {Poggianti}, {Moretti},
  {Gullieuszik}, {Smith}, {Vulcani}, {Fasano}, {Fritz}, {Tonnesen}, {Bettoni},
  {Hau}, {Biviano}, {Bellhouse}, \& {McGee}}]{gaspIXerr}
---. 2019, \mnras, 482, 3454

\bibitem[{{Kapferer} {et~al.}(2009){Kapferer}, {Sluka}, {Schindler}, {Ferrari},
  \& {Ziegler}}]{kapf+2009}
{Kapferer}, W., {Sluka}, C., {Schindler}, S., {Ferrari}, C., \& {Ziegler}, B.
  2009, \aap, 499, 87

\bibitem[{{Kauffmann} {et~al.}(2003){Kauffmann}, {Heckman}, {Tremonti},
  {Brinchmann}, {Charlot}, {White}, {Ridgway}, {Brinkmann}, {Fukugita}, {Hall},
  {Ivezi{\'c}}, {Richards}, \& {Schneider}}]{kauf+2003}
{Kauffmann}, G., {Heckman}, T.~M., {Tremonti}, C., {et~al.} 2003, \mnras, 346,
  1055

\bibitem[{{Kenney} {et~al.}(2014){Kenney}, {Geha}, {J{\'a}chym}, {Crowl},
  {Dague}, {Chung}, {van Gorkom}, \& {Vollmer}}]{kenn+2014}
{Kenney}, J.~D.~P., {Geha}, M., {J{\'a}chym}, P., {et~al.} 2014, \apj, 780, 119

\bibitem[{{Kenney} {et~al.}(2004){Kenney}, {van Gorkom}, \&
  {Vollmer}}]{kenn+2004}
{Kenney}, J.~D.~P., {van Gorkom}, J.~H., \& {Vollmer}, B. 2004, \aj, 127, 3361

\bibitem[{{Kewley} {et~al.}(2001){Kewley}, {Dopita}, {Sutherland}, {Heisler},
  \& {Trevena}}]{kewl+2001}
{Kewley}, L.~J., {Dopita}, M.~A., {Sutherland}, R.~S., {Heisler}, C.~A., \&
  {Trevena}, J. 2001, \apj, 556, 121

\bibitem[{{Leauthaud} {et~al.}(2010){Leauthaud}, {Finoguenov}, {Kneib},
  {Taylor}, {Massey}, {Rhodes}, {Ilbert}, {Bundy}, {Tinker}, {George}, {Capak},
  {Koekemoer}, \& et~al.}]{leauthaud+2010}
{Leauthaud}, A., {Finoguenov}, A., {Kneib}, J.-P., {et~al.} 2010, \apj, 709, 97

\bibitem[{{Leroy} {et~al.}(2008){Leroy}, {Walter}, {Brinks}, {Bigiel}, {de
  Blok}, {Madore}, \& {Thornley}}]{lero+2008}
{Leroy}, A.~K., {Walter}, F., {Brinks}, E., {et~al.} 2008, \aj, 136, 2782

\bibitem[{{McPartland} {et~al.}(2016){McPartland}, {Ebeling}, {Roediger}, \&
  {Blumenthal}}]{mcpa+2016}
{McPartland}, C., {Ebeling}, H., {Roediger}, E., \& {Blumenthal}, K. 2016,
  \mnras, 455, 2994

\bibitem[{{Melnick} {et~al.}(2012){Melnick}, {Giraud}, {Toledo}, {Selman}, \&
  {Quintana}}]{meln+2012}
{Melnick}, J., {Giraud}, E., {Toledo}, I., {Selman}, F., \& {Quintana}, H.
  2012, \mnras, 427, 850

\bibitem[{{Merluzzi} {et~al.}(2013){Merluzzi}, {Busarello}, {Dopita}, {Haines},
  {Steinhauser}, {Mercurio}, {Rifatto}, {Smith}, \& {Schindler}}]{merl+2013}
{Merluzzi}, P., {Busarello}, G., {Dopita}, M.~A., {et~al.} 2013, \mnras, 429,
  1747

\bibitem[{{Montes} \& {Trujillo}(2018)}]{mont+2018}
{Montes}, M., \& {Trujillo}, I. 2018, \mnras, 474, 917

\bibitem[{{Moretti} {et~al.}(2017){Moretti}, {Gullieuszik}, {Poggianti},
  {Paccagnella}, {Couch}, {Vulcani}, {Bettoni}, {Fritz}, {Cava}, {Fasano},
  {D'Onofrio}, \& {Omizzolo}}]{more+2017}
{Moretti}, A., {Gullieuszik}, M., {Poggianti}, B., {et~al.} 2017, \aap, 599,
  A81

\bibitem[{{Moretti} {et~al.}(2018{\natexlab{a}}){Moretti}, {Paladino},
  {Poggianti}, {D'Onofrio}, {Bettoni}, {Gullieuszik}, {Jaff{\'e}}, {Vulcani},
  {Fasano}, {Fritz}, \& {Torstensson}}]{more+2018}
{Moretti}, A., {Paladino}, R., {Poggianti}, B.~M., {et~al.} 2018{\natexlab{a}},
  \mnras, 480, 2508

\bibitem[{{Moretti} {et~al.}(2018{\natexlab{b}}){Moretti}, {Paladino},
  {Poggianti}, {D'Onofrio}, {Bettoni}, {Gullieuszik}, {Jaff{\'e}}, {Vulcani},
  {Fasano}, {Fritz}, \& {Torstensson}}]{gaspX}
---. 2018{\natexlab{b}}, \mnras, 480, 2508

\bibitem[{{Moretti} {et~al.}(2018{\natexlab{c}}){Moretti}, {Poggianti},
  {Gullieuszik}, {Mapelli}, {Jaff{\'e}}, {Fritz}, {Biviano}, {Fasano},
  {Bettoni}, {Vulcani}, \& {D'Onofrio}}]{gaspV}
{Moretti}, A., {Poggianti}, B.~M., {Gullieuszik}, M., {et~al.}
  2018{\natexlab{c}}, \mnras, 475, 4055

\bibitem[{{Moretti} {et~al.}(2020{\natexlab{a}}){Moretti}, {Paladino},
  {Poggianti}, {Serra}, {Roediger}, {Gullieuszik}, {Tomi{\v{c}}i{\'c}},
  {Radovich}, {Vulcani}, {Jaff{\'e}}, {Fritz}, {Bettoni}, {Ramatsoku}, \&
  {Wolter}}]{more+2020}
{Moretti}, A., {Paladino}, R., {Poggianti}, B.~M., {et~al.} 2020{\natexlab{a}},
  \apj, 889, 9

\bibitem[{{Moretti} {et~al.}(2020{\natexlab{b}}){Moretti}, {Paladino},
  {Poggianti}, {Serra}, {Ramatsoku}, {Franchetto}, {Deb}, {Gullieuszik},
  {Tomicic}, {Mingozzi}, {Vulcani}, {Radovich}, {Bettoni}, \&
  {Fritz}}]{more+2020ALMA}
---. 2020{\natexlab{b}}, arXiv e-prints, arXiv:2006.13612

\bibitem[{{Morishita} {et~al.}(2017){Morishita}, {Abramson}, {Treu}, {Schmidt},
  {Vulcani}, \& {Wang}}]{mori+2017}
{Morishita}, T., {Abramson}, L.~E., {Treu}, T., {et~al.} 2017, \apj, 846, 139

\bibitem[{{Munari} {et~al.}(2013){Munari}, {Biviano}, {Borgani}, {Murante}, \&
  {Fabjan}}]{muna+2013}
{Munari}, E., {Biviano}, A., {Borgani}, S., {Murante}, G., \& {Fabjan}, D.
  2013, \mnras, 430, 2638

\bibitem[{{Owers} {et~al.}(2019){Owers}, {Hudson}, {Oman}, {Bland -Hawthorn},
  {Brough}, {Bryant}, {Cortese}, {Couch}, {Croom}, {van de Sande}, {Federrath},
  {Groves}, {Hopkins}, {Lawrence}, {Lorente}, {McDermid}, {Medling},
  {Richards}, {Scott}, {Taranu}, {Welker}, \& {Yi}}]{ower+2019}
{Owers}, M.~S., {Hudson}, M.~J., {Oman}, K.~A., {et~al.} 2019, \apj, 873, 52

\bibitem[{{Planck Collaboration} {et~al.}(2014){Planck Collaboration}, {Ade},
  {Aghanim}, {Armitage-Caplan}, {Arnaud}, {Ashdown}, {Atrio-Barandela},
  {Aumont}, {Baccigalupi}, {Banday}, \& et~al.}]{Planck_XVI+2014}
{Planck Collaboration}, {Ade}, P.~A.~R., {Aghanim}, N., {et~al.} 2014, \aap,
  571, A16

\bibitem[{{Poggianti} {et~al.}(2016){Poggianti}, {Fasano}, {Omizzolo},
  {Gullieuszik}, {Bettoni}, {Moretti}, {Paccagnella}, {Jaff{\'e}}, {Vulcani},
  {Fritz}, {Couch}, \& {D'Onofrio}}]{pogg+2016}
{Poggianti}, B.~M., {Fasano}, G., {Omizzolo}, A., {et~al.} 2016, \aj, 151, 78

\bibitem[{{Poggianti} {et~al.}(2017{\natexlab{a}}){Poggianti}, {Moretti},
  {Gullieuszik}, {Fritz}, {Jaff{\'e}}, {Bettoni}, {Fasano}, {Bellhouse}, {Hau},
  {Vulcani}, {Biviano}, {Omizzolo}, {Paccagnella}, {D'Onofrio}, {Cava},
  {Sheen}, {Couch}, \& {Owers}}]{gaspI}
{Poggianti}, B.~M., {Moretti}, A., {Gullieuszik}, M., {et~al.}
  2017{\natexlab{a}}, \apj, 844, 48

\bibitem[{{Poggianti} {et~al.}(2017{\natexlab{b}}){Poggianti}, {Jaff{\'e}},
  {Moretti}, {Gullieuszik}, {Radovich}, {Tonnesen}, {Fritz}, {Bettoni},
  {Vulcani}, {Fasano}, {Bellhouse}, {Hau}, \& {Omizzolo}}]{gaspVI}
{Poggianti}, B.~M., {Jaff{\'e}}, Y.~L., {Moretti}, A., {et~al.}
  2017{\natexlab{b}}, \nat, 548, 304

\bibitem[{{Poggianti} {et~al.}(2019{\natexlab{a}}){Poggianti}, {Gullieuszik},
  {Tonnesen}, {Moretti}, {Vulcani}, {Radovich}, {Jaff{\'e}}, {Fritz},
  {Bettoni}, {Franchetto}, {Fasano}, {Bellhouse}, \& {Omizzolo}}]{gaspXIII}
{Poggianti}, B.~M., {Gullieuszik}, M., {Tonnesen}, S., {et~al.}
  2019{\natexlab{a}}, \mnras, 482, 4466

\bibitem[{{Poggianti} {et~al.}(2019{\natexlab{b}}){Poggianti}, {Ignesti},
  {Gitti}, {Wolter}, {Brighenti}, {Biviano}, {George}, {Vulcani},
  {Gullieuszik}, {Moretti}, {Paladino}, {Bettoni}, {Franchetto}, {Jaff{\'e}},
  {Radovich}, {Roediger}, {Tomi{\v{c}}i{\'c}}, {Tonnesen}, {Bellhouse},
  {Fritz}, \& {Omizzolo}}]{pogg+jw100}
{Poggianti}, B.~M., {Ignesti}, A., {Gitti}, M., {et~al.} 2019{\natexlab{b}},
  \apj, 887, 155

\bibitem[{{Popping} {et~al.}(2014){Popping}, {Somerville}, \&
  {Trager}}]{popp+2014}
{Popping}, G., {Somerville}, R.~S., \& {Trager}, S.~C. 2014, \mnras, 442, 2398

\bibitem[{{Radovich} {et~al.}(2019){Radovich}, {Poggianti}, {Jaff{\'e}},
  {Moretti}, {Bettoni}, {Gullieuszik}, {Vulcani}, \& {Fritz}}]{gaspXIX}
{Radovich}, M., {Poggianti}, B., {Jaff{\'e}}, Y.~L., {et~al.} 2019, \mnras,
  486, 486

\bibitem[{{Ramatsoku} {et~al.}(2019){Ramatsoku}, {Serra}, {Poggianti},
  {Moretti}, {Gullieuszik}, {Bettoni}, {Deb}, {Fritz}, {van Gorkom},
  {Jaff{\'e}}, {Tonnesen}, {Verheijen}, {Vulcani}, {Hugo}, {J{\'o}zsa},
  {Maccagni}, {Makhathini}, {Ramaila}, {Smirnov}, \& {Thorat}}]{gaspXVII}
{Ramatsoku}, M., {Serra}, P., {Poggianti}, B.~M., {et~al.} 2019, \mnras, 487,
  4580

\bibitem[{{Roediger} {et~al.}(2014){Roediger}, {Bruggen}, {Owers}, {Ebeling},
  \& {Sun}}]{roed+2014}
{Roediger}, E., {Bruggen}, M., {Owers}, M.~S., {Ebeling}, H., \& {Sun}, M.
  2014, \mnras, 443, L114

\bibitem[{{Schlafly} \& {Finkbeiner}(2011)}]{schl+2011}
{Schlafly}, E.~F., \& {Finkbeiner}, D.~P. 2011, \apj, 737, 103

\bibitem[{{Sharp} \& {Bland-Hawthorn}(2010)}]{shar+2010}
{Sharp}, R.~G., \& {Bland-Hawthorn}, J. 2010, \apj, 711, 818

\bibitem[{{Smith} {et~al.}(2012){Smith}, {Fellhauer}, \& {Assmann}}]{smit+2012}
{Smith}, R., {Fellhauer}, M., \& {Assmann}, P. 2012, \mnras, 420, 1990

\bibitem[{{Smith} {et~al.}(2010){Smith}, {Lucey}, {Hammer}, {Hornschemeier},
  {Carter}, {Hudson}, {Marzke}, {Mouhcine}, {Eftekharzadeh}, {James},
  {Khosroshahi}, {Kourkchi}, \& {Karick}}]{smit+2010}
{Smith}, R.~J., {Lucey}, J.~R., {Hammer}, D., {et~al.} 2010, \mnras, 408, 1417

\bibitem[{{Springel} {et~al.}(2005){Springel}, {White}, {Jenkins}, {Frenk},
  {Yoshida}, {Gao}, {Navarro}, {Thacker}, {Croton}, {Helly}, {Peacock}, {Cole},
  {Thomas}, {Couchman}, {Evrard}, {Colberg}, \& {Pearce}}]{Springel+2005}
{Springel}, V., {White}, S.~D.~M., {Jenkins}, A., {et~al.} 2005, \nat, 435, 629

\bibitem[{{Sun} {et~al.}(2006){Sun}, {Jones}, {Forman}, {Nulsen}, {Donahue}, \&
  {Voit}}]{sun+2006}
{Sun}, M., {Jones}, C., {Forman}, W., {et~al.} 2006, \apjl, 637, L81

\bibitem[{{The Astropy Collaboration} {et~al.}(2018){The Astropy
  Collaboration}, {Price-Whelan}, {Sip{\H o}cz}, {G{\"u}nther}, {Lim},
  {Crawford}, {Conseil}, {Shupe}, {Craig}, {Dencheva}, {Ginsburg},
  {VanderPlas}, {Bradley}, {P{\'e}rez-Su{\'a}rez}, {de Val-Borro}, {Aldcroft},
  {Cruz}, {Robitaille}, {Tollerud}, {Ardelean}, {Babej}, {Bachetti}, {Bakanov},
  {Bamford}, {Barentsen}, {Barmby}, {Baumbach}, {Berry}, {Biscani}, {Boquien},
  {Bostroem}, {Bouma}, {Brammer}, {Bray}, {Breytenbach}, {Buddelmeijer},
  {Burke}, {Calderone}, {Cano Rodr{\'{\i}}guez}, {Cara}, {Cardoso},
  {Cheedella}, {Copin}, {Crichton}, {D{\'A}vella}, {Deil}, {Depagne},
  {Dietrich}, {Donath}, {Droettboom}, {Earl}, {Erben}, {Fabbro}, {Ferreira},
  {Finethy}, {Fox}, {Garrison}, {Gibbons}, {Goldstein}, {Gommers}, {Greco},
  {Greenfield}, {Groener}, {Grollier}, {Hagen}, {Hirst}, {Homeier}, {Horton},
  {Hosseinzadeh}, {Hu}, {Hunkeler}, {Ivezi{\'c}}, {Jain}, {Jenness}, {Kanarek},
  {Kendrew}, {Kern}, {Kerzendorf}, {Khvalko}, {King}, {Kirkby}, {Kulkarni},
  {Kumar}, {Lee}, {Lenz}, {Littlefair}, {Ma}, {Macleod}, {Mastropietro},
  {McCully}, {Montagnac}, {Morris}, {Mueller}, {Mumford}, {Muna}, {Murphy},
  {Nelson}, {Nguyen}, {Ninan}, {N{\"o}the}, {Ogaz}, {Oh}, {Parejko}, {Parley},
  {Pascual}, {Patil}, {Patil}, {Plunkett}, {Prochaska}, {Rastogi}, {Reddy
  Janga}, {Sabater}, {Sakurikar}, {Seifert}, {Sherbert}, {Sherwood-Taylor},
  {Shih}, {Sick}, {Silbiger}, {Singanamalla}, {Singer}, {Sladen}, {Sooley},
  {Sornarajah}, {Streicher}, {Teuben}, {Thomas}, {Tremblay}, {Turner},
  {Terr{\'o}n}, {van Kerkwijk}, {de la Vega}, {Watkins}, {Weaver}, {Whitmore},
  {Woillez}, \& {Zabalza}}]{astropy}
{The Astropy Collaboration}, {Price-Whelan}, A.~M., {Sip{\H o}cz}, B.~M.,
  {et~al.} 2018, ArXiv e-prints, arXiv:1801.02634

\bibitem[{{Tonnesen} \& {Bryan}(2012)}]{tonn+2012}
{Tonnesen}, S., \& {Bryan}, G.~L. 2012, \mnras, 422, 1609

\bibitem[{{Verdugo} {et~al.}(2015){Verdugo}, {Combes}, {Dasyra}, {Salom{\'e}},
  \& {Braine}}]{verd+2015}
{Verdugo}, C., {Combes}, F., {Dasyra}, K., {Salom{\'e}}, P., \& {Braine}, J.
  2015, \aap, 582, A6

\bibitem[{{Vollmer} {et~al.}(2009){Vollmer}, {Soida}, {Chung}, {Chemin},
  {Braine}, {Boselli}, \& {Beck}}]{voll+2009}
{Vollmer}, B., {Soida}, M., {Chung}, A., {et~al.} 2009, \aap, 496, 669

\bibitem[{{Vulcani} {et~al.}(2018){Vulcani}, {Poggianti}, {Gullieuszik},
  {Moretti}, {Tonnesen}, {Jaff{\'e}}, {Fritz}, {Fasano}, \&
  {Bettoni}}]{gaspXIV}
{Vulcani}, B., {Poggianti}, B.~M., {Gullieuszik}, M., {et~al.} 2018, \apjl,
  866, L25

\bibitem[{{Wu}(2018)}]{wu2018}
{Wu}, P.-F. 2018, \mnras, 473, 5468

\bibitem[{{Yagi} {et~al.}(2010){Yagi}, {Yoshida}, {Komiyama}, {Kashikawa},
  {Furusawa}, {Okamura}, {Graham}, {Miller}, {Carter}, {Mobasher}, \&
  {Jogee}}]{yagi+2010}
{Yagi}, M., {Yoshida}, M., {Komiyama}, Y., {et~al.} 2010, \aj, 140, 1814

\bibitem[{{Yim} {et~al.}(2014){Yim}, {Wong}, {Xue}, {Rand}, {Rosolowsky}, {van
  der Hulst}, {Benjamin}, \& {Murphy}}]{yim+2014}
{Yim}, K., {Wong}, T., {Xue}, R., {et~al.} 2014, \aj, 148, 127

\bibitem[{{Yoon} {et~al.}(2017){Yoon}, {Chung}, {Smith}, \&
  {Jaff{\'e}}}]{yoon+2017}
{Yoon}, H., {Chung}, A., {Smith}, R., \& {Jaff{\'e}}, Y.~L. 2017, \apj, 838, 81

\bibitem[{{Yoshida} {et~al.}(2004){Yoshida}, {Ohyama}, {Iye}, {Aoki},
  {Kashikawa}, {Sasaki}, {Shimasaku}, {Yagi}, {Okamura}, {Doi}, {Furusawa},
  {Hamabe}, {Kimura}, {Komiyama}, {Miyazaki}, {Miyazaki}, {Nakata}, {Ouchi},
  {Sekiguchi}, \& {Yasuda}}]{yosh+2004}
{Yoshida}, M., {Ohyama}, Y., {Iye}, M., {et~al.} 2004, \aj, 127, 90

\end{thebibliography}

\appendix
\section{Truncation radius}

\begin{figure}
\centering
\includegraphics[width=.46\textwidth]{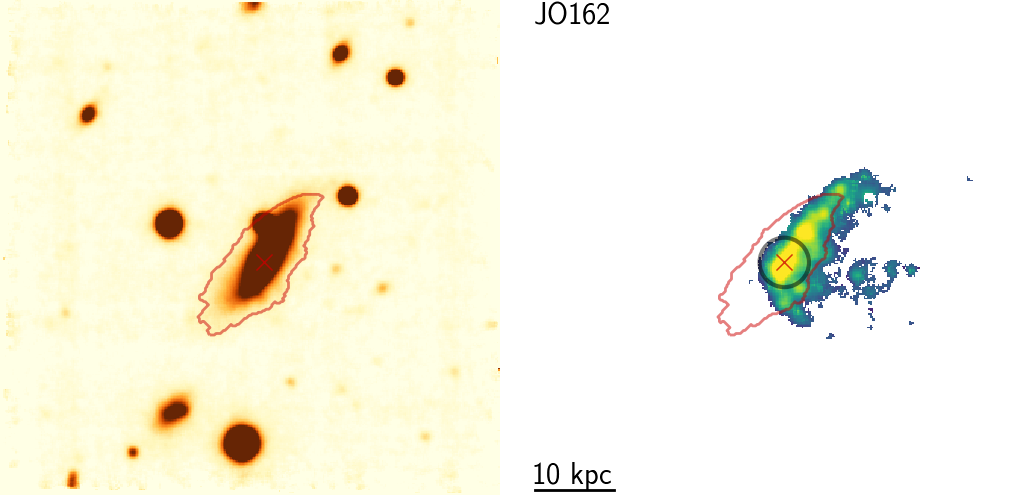}\hfill
\includegraphics[width=.46\textwidth]{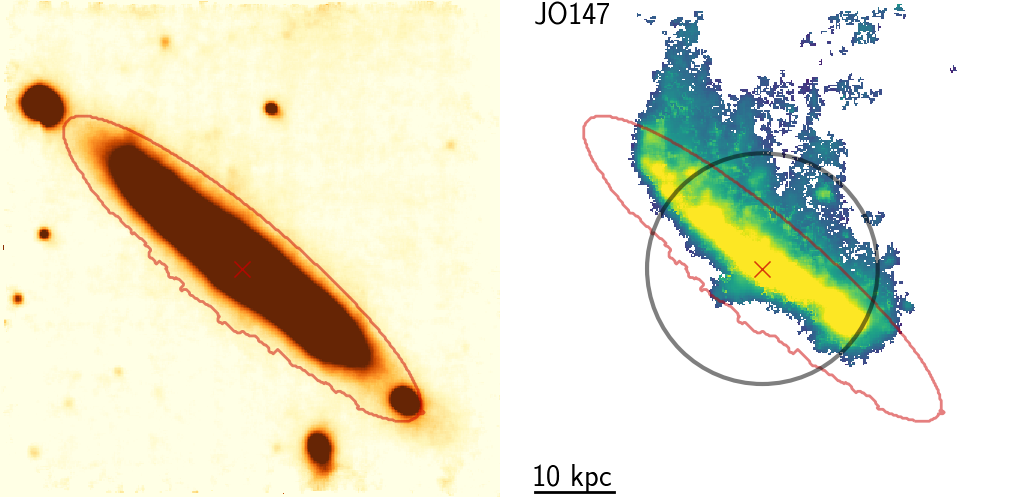}\\[5pt]
\includegraphics[width=.46\textwidth]{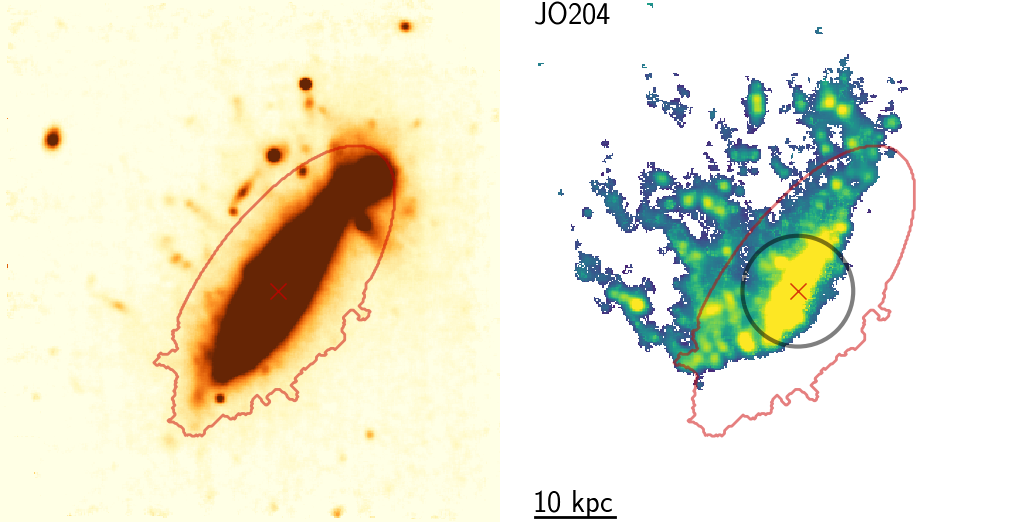}\hfill
\includegraphics[width=.46\textwidth]{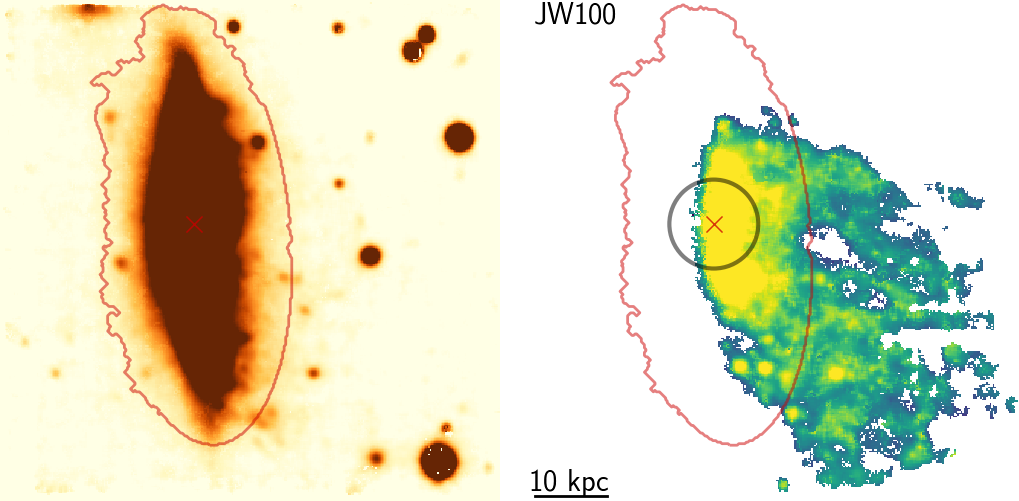}\\[5pt]
\includegraphics[width=.94\textwidth]{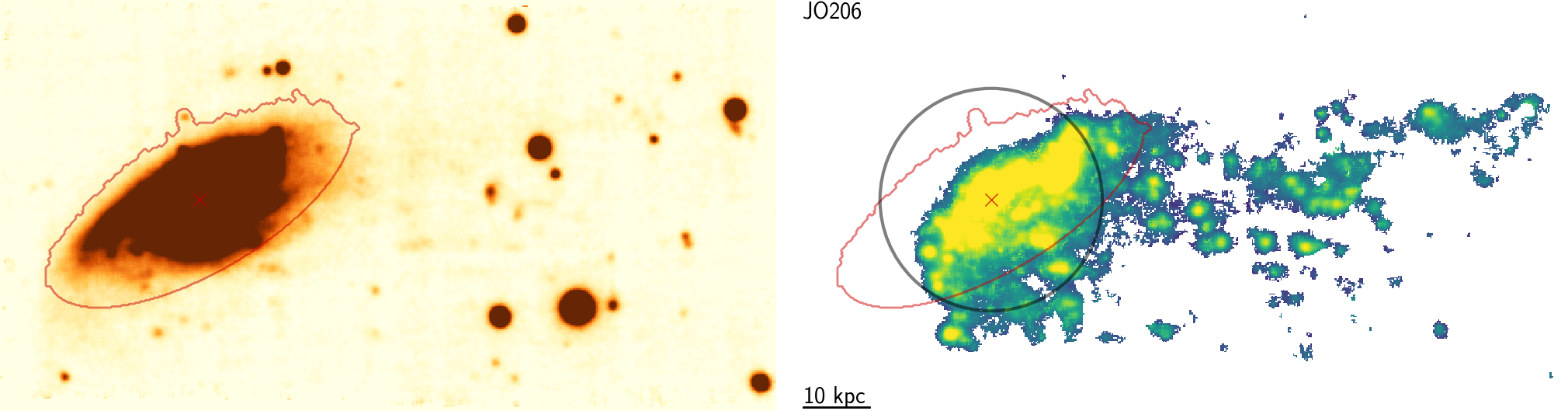}
\caption{Continuum emission map, H$\alpha$ emission, and the contour used to define the galaxy main body as in Fig.~\ref{fig:inoutmap} for jellyfish galaxies with extended tails. The blue circle show the truncation radius $r_t$ computed using the model described in Sect. \ref{sec:model}.}
\label{fig:rt}
\end{figure}

Figure~\ref{fig:rt} shows examples of the the computed $r_t$  compared with the H$\alpha$ distribution. Only galaxies with extended tails were considered, excluding cases of edge-on stripping.

\facilities{VLT (MUSE)}

\software{KUBEVIZ, SINOPSIS, IDL, Python, astropy.}
\end{document}